\documentclass[twocolumn,aps,pre,amsmath,showpacs]{revtex4}
\usepackage{graphicx}

\begin{document}

\title{Structural relaxation in supercooled orthoterphenyl}
\author{S.-H. Chong\footnote{Present address: 
Laboratoire de Physique Math\'ematique et Th\'eorique,
Universit\'e Montpellier II, 
34095 Montpellier, France}
and F. Sciortino}
\affiliation{Dipartimento di Fisica and
Istituto Nazionale per la Fisica della Materia,
Center for Statistical Mechanics and Complexity,
Universit\`a di Roma ``La Sapienza,''
Piazzale Aldo Moro 2, I-00185, Roma, Italy}
\date{\today}

\begin{abstract}

We report molecular-dynamics simulation results 
performed for a model of 
molecular liquid orthoterphenyl in supercooled states,
which we then compare with both experimental data
and mode-coupling-theory (MCT) predictions,
aiming at a better understanding of 
structural relaxation in orthoterphenyl. 
We pay special attention to the wavenumber dependence 
of the collective dynamics. 
It is shown that the simulation results for the model
share many features with experimental data for real system, 
and that MCT captures the simulation results
at the semiquantitative level
except for intermediate wavenumbers connected to the overall
size of the molecule. 
Theoretical results at the intermediate wavenumber region
are found to be improved by taking into account the spatial
correlation of the molecule's geometrical center.
This supports the idea that unusual dynamical properties at 
the intermediate wavenumbers, reported previously 
in simulation studies for the model
and discernible in coherent neutron-scattering experimental data, 
are basically due to the coupling of the rotational motion to the
geometrical-center dynamics. 
However, there still remain qualitative as well as quantitative 
discrepancies between theoretical prediction and corresponding 
simulation results at the intermediate wavenumbers,
which call for further theoretical investigation.

\end{abstract}

\pacs{61.20.Ja,61.20.Lc,61.25.Em,64.70.Pf}

\maketitle

\section{Introduction}

Describing the microscopic origin of structural slowing down
on cooling or compressing glass-forming liquids is one of the most
challenging problems in condensed matter physics.
During the past decade the research in this field was strongly influenced 
by the idealized mode-coupling theory (MCT) for the evolution
of structural relaxation~\cite{Goetze91b,Goetze92,Goetze99}. 
The theory predicts the structural arrest --
also referred to as the idealized liquid-glass transition --
driven by the mutual blocking of a particle and its neighbors
(``cage effect'')
at a critical temperature $T_{c}$ which is located above the 
glass-transition temperature $T_{g}$. 
For temperatures close to but above $T_{c}$, 
MCT predicts universal scaling laws and
power laws for describing the glassy slow structural relaxation. 
Although  such complete structural arrest at $T_{c}$
is not observed in experiments and therefore the idealized theory cannot
literally be applied for describing dynamics below $T_{c}$, 
extensive tests of the theoretical predictions carried out so far
against experimental and computer-simulation results 
suggest that MCT deals properly with some essential features of 
supercooled liquids~\cite{Goetze92,Goetze99}. 

The molecular van der Waals liquid orthoterphenyl (OTP)
has long been used as a model system in the study of the
glass transition.
(See, e.g., Refs.~\cite{Toelle97,Toelle01,Monaco01} and papers quoted therein.) 
Extensive experiments on OTP
have been performed to monitor the onset 
of glassy structural relaxation on microscopic time and length scales.
Using in particular quasielastic neutron scattering,
the decay of collective and self density fluctuations
has been measured as a function of temperature, pressure, and
wavenumber~\cite{Toelle97,Petry91,Kiebel92,Wuttke93,Bartsch95,Toelle98b,Toelle98}. 
Based on these studies, 
the validity of the {\em universal} predictions of MCT,
such as the factorization theorem and the time-temperature superposition principle,
has been established.
However, there are {\em nonuniversal} aspects  
in the glassy structural relaxation
which cannot be elucidated solely by those universal predictions. 
For example, 
parameters describing the decay of the collective density fluctuations
in the $\alpha$-relaxation regime
exhibit a characteristic wavenumber dependence, which is an
important nonuniversal aspect 
to be accounted for by the theory~\cite{Toelle97,Toelle98}. 
As will be discussed below, 
we found from the analysis of molecular-dynamics (MD) simulation results
interesting dynamical properties of OTP at intermediate
wavenumbers, and their theoretical investigation
is one of the main points of the present paper. 

One of the distinctive features of MCT is that it can also  
make predictions concerning nonuniversal aspects, such as 
the value of $T_{c}$ and the details of the 
time and wavenumber dependence of various dynamical quantities, 
provided the system's static structure factor is known with 
sufficient accuracy.
Utilizing this feature, 
there have been quantitative
tests of the theory concerning nonuniversal as well as universal aspects
using as input only the static structure factor
determined from integral-equation theories or
from computer simulations~\cite{Goetze91,Foffi03,Nauroth97,Sciortino01,Winkler00,Fabbian99b,Theis00}.
In the present paper, we apply MCT to discuss in detail
properties of the slow structural relaxation in OTP,
paying special attention to the wavenumber dependence
of the structural $\alpha$ relaxation of the collective dynamics. 
This will done by 
regarding MD simulation results as a bridge connecting
experimental data and theoretical predictions. 

Although OTP is one of the simplest molecular systems from
an experimental side, it is rather a complicated
system for a theoretical treatment.
Therefore, it is unavoidable to deal with a simple model for OTP
which is still efficient in mimicking the complexity of the dynamical
behavior of real system.
In this respect, 
Lewis and Wahnstr\"om (LW)~\cite{Lewis94} introduced a particularly useful 
three-site model, 
each site playing the role of a whole phenyl ring,
and this model will be considered in this paper.

For the LW OTP model, 
static and dynamic properties in supercooled states have been extensively studied 
based on MD simulations~\cite{Rinaldi01,Mossa02,Chong03}.
In the present work, the simulation results for 
the static structure factors 
serve as input to the theory, and those for dynamics can be 
used in testing the so-obtained theoretical results. 
Thereby, a stringent test of MCT predictions against the simulation 
results can be performed. 
Furthermore, by making connections between the simulation results for LW OTP
and experimental ones for real system, 
the relevance of our theoretical results and their interpretation
in understanding experimental data can be established. 

A theoretical analysis for LW OTP has already been presented in 
Ref.~\cite{Rinaldi01} based on a simplified theory which is essentially the
same as the one for spherical particles. 
But, since OTP is a molecular system, a full molecular approach 
is desirable.
Recently, MCT for spherical particles has been extended to a theory for
molecules. 
This has been done based on the tensor- and site-density
formulations. 
In the tensor-density formulation, the density-fluctuation correlator is 
generalized to the one of infinite matrices of
correlation functions formed with tensor-density 
fluctuations~\cite{Winkler00,Fabbian99b,Theis00,Schilling97,Theis98,Letz00,Theenhaus01,Franosch97c,Goetze00c}.
However, the tensor-density formulation has the difficulty that the resulting
equations are so involved,
and it is not obvious whether those equations can be numerically
solved within the regime of glassy dynamics. 
To overcome this difficulty, 
it has been suggested to base MCT for molecular 
systems on the site 
representation~\cite{Chong98b,Chong01,Chong01b,Chong02,Chong02b}. 
The fluctuations of the interaction-site densities
have been used as the basic variables to describe
the structure of the system.
As a result, the known scalar MCT equations for the density
fluctuations in simple systems have been generalized to 
$n$-by-$n$ matrix equations for the interaction-site-density
fluctuations, where $n$ denotes the number of atoms (or sites)
forming the molecule.
Thus, relatively simple equations of motions can be obtained
within the site-representation, and these MCT equations 
will be solved in the present work 
to discuss the structural relaxation in OTP.

The paper is organized as follows.
In Sec.~\ref{sec:model}, the LW OTP model shall be introduced, and
static as well as dynamic quantities to be used in discussions
of simulation and theoretical results are defined. 
In Sec.~\ref{sec:MD}, MD-simulation results for LW OTP
are summarized, and possible connections to the experimental results 
for real system are established. 
In Sec.~\ref{sec:MCT}, 
after reviewing relevant MCT equations based on the site
representation, theoretical results are
presented and compared with the simulation results. 
The paper is summarized in Sec.~\ref{sec:conclusions} with
some concluding remarks.

\section{Model}
\label{sec:model}

\begin{figure}
\includegraphics[width=0.8\linewidth]{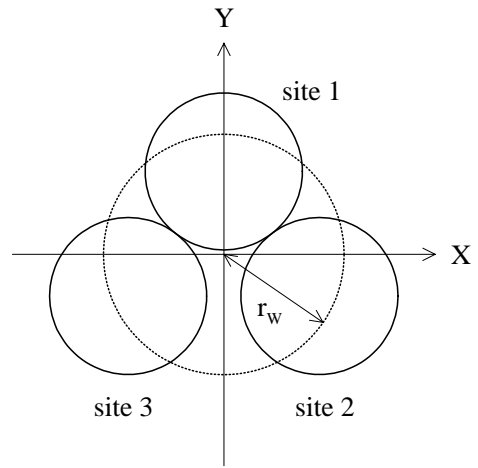}
\caption{
Schematic representation of the three-site LW OTP molecule
in the body-fixed molecular
XY plane where the origin is taken to be the geometrical
center (GC),
${\vec r}_{j}^{\, \rm GC} = (1/3) \sum_{a=1}^{3} {\vec r}_{j}^{\, a}$, 
and the Y axis is along the symmetry axis of
the molecule.
The diameter of each site is taken from the
LJ parameter $\sigma = 0.483$ nm.
The dotted circle is drawn with the van der Waals
radius $r_{\rm W} = 0.37$ nm~\protect\cite{Bondi64}
and taking GC as the origin.}
\label{fig:picture-OTP}
\end{figure}

The LW OTP molecule 
designed by Lewis and Wahnstr\"om~\cite{Lewis94}
is a rigid isosceles triangle;
the length of the two short sides of the triangle is 
$\ell = 0.483$ nm and the angle between them is $\theta = 75^{\circ}$.
Each of the three sites represents an entire phenyl ring of 
mass $m \approx 78$ amu,
and is described by a Lennard-Jones (LJ) sphere whose interaction
potential is given by
\begin{equation}
V(r) = 4 \epsilon \, 
[(\sigma/r)^{12} - (\sigma/r)^{6}] + A + Br,
\end{equation}
with $\epsilon = 5.276$ kJ/mol,
$\sigma = 0.483$ nm,
$A = 0.461$ kJ/mol, and
$B = -0.313$ kJ/(mol nm).
Schematic representation of the LW OTP molecule is presented
in Fig.~\ref{fig:picture-OTP}.
The shape of the molecule and the 
LJ parameters have been chosen to reproduce some bulk properties 
of the OTP molecule. 
The values of $A$ and $B$ are selected in such a way that
the potential and its first derivative are zero at the cutoff
$r_{c} = 1.2616$ nm adopted in MD simulations.
MD simulation results on the thermodynamic and dynamic properties 
of LW OTP have been reported in Refs.~\cite{Lewis94,Rinaldi01,Mossa02,Chong03}. 

Discussions on the dynamics shall be done 
based on the site-site density correlators
\begin{equation}
F_{q}^{ab}(t) = 
\langle \rho_{\vec q}^{a}(t)^{*} \rho_{\vec q}^{b}(0) \rangle / N \quad
(a,b = 1,2,3),
\end{equation}
and their derivatives to be defined below. 
Here
$\rho_{\vec q}^{a}(t) = \sum_{j=1}^{N}
\exp[i {\vec q} \cdot {\vec r}_{j}^{\, a}(t)]$,
with ${\vec r}_{j}^{\, a}(t)$ being the position vector of site $a$ 
in $j$th molecule at time $t$, 
denotes the site-density fluctuations, 
$N$ the number of molecules in the system,
and $\langle \cdots \rangle$ canonical averaging for temperature $T$. 
(As depicted in Fig.~\ref{fig:picture-OTP}, 
we shall use the convention that $a=1$ refers to the central site 
and $a=2,3$ to the adjacent sites.)
Because of isotropy, $F_{q}^{ab}(t)$ depends
only on the wavenumber $q = | \, {\vec q} \, |$.
The initial values constitute the site-site static
structure factors, $S_{q}^{ab} = F_{q}^{ab}(0)$,
which provide the simplest information on the equilibrium
structure of the system. 

We also introduce tagged-molecule correlators, 
$F_{q,s}^{ab}(t) = 
\langle \rho_{{\vec q},s}^{a}(t)^{*} \rho_{{\vec q},s}^{b}(0) \rangle$,
in which
$\rho_{{\vec q},s}^{a}(t) = \exp[i {\vec q} \cdot {\vec r}_{s}^{\, a}(t)]$
with ${\vec r}_{s}^{\, a}(t)$ denoting the position vector of site $a$
in the tagged molecule (labeled $s$) at time $t$.
The initial value shall be denoted as 
$w_{q}^{ab} = F_{q,s}^{ab}(0)$.
For a rigid molecule, 
it is given by 
$w_{q}^{ab} = \delta^{ab} + (1 - \delta^{ab}) j_{0}(q l^{ab})$, 
where $j_{0}(x)$ denotes the 0th-order spherical Bessel function
and $l^{ab}$ the distance between sites $a$ and $b$.

Let us consider a more convenient representation of the site-site
density correlators which exploits the $C_{2v}$ symmetry of the
LW OTP molecule. 
For this purpose, we introduce the following density correlators
\begin{equation}
F_{q}^{\rm X}(t) = 
\langle \rho_{\vec q}^{\rm X}(t)^{*} \rho_{\vec q}^{\rm X}(0) \rangle / N
\,\, \mbox{ for } \,\, 
\mbox{X = N, Z, and Q},
\end{equation}
defined in terms of the linear combinations
\begin{eqnarray}
& &
\rho_{\vec q}^{\rm N} = 
(\rho_{\vec q}^{1} + \rho_{\vec q}^{2} + \rho_{\vec q}^{3}) / \sqrt{3}, \quad 
\rho_{\vec q}^{\rm Z} = 
(2\rho_{\vec q}^{1} - \rho_{\vec q}^{2} - \rho_{\vec q}^{3}) / \sqrt{6}, 
\nonumber \\
& &
\qquad \qquad \qquad \quad
\rho_{\vec q}^{\rm Q} = 
(\rho_{\vec q}^{2} - \rho_{\vec q}^{3}) / \sqrt{2},
\end{eqnarray}
of the site-density fluctuations
$\rho_{\vec q}^{a}$ ($a=1$, 2, or 3).
One can easily show that $F_{q}^{\rm X}(t)$ can be expressed
in terms of $F_{q}^{ab}(t)$: one finds, for example, 
$F_{q}^{\rm N}(t) = \sum_{a,b=1}^{3} F_{q}^{ab}(t) / 3$.
Their normalized correlators shall be denoted as
$\phi_{q}^{\rm X}(t) = F_{q}^{\rm X}(t) / S_{q}^{\rm X}$
with the corresponding static structure factor
$S_{q}^{\rm X} = F_{q}^{\rm X}(0)$. 
We also introduce self-parts of these correlators, 
to be denoted as 
$F_{q,s}^{\rm X}(t)$, 
which are defined similarly in terms of the site-density fluctuations
$\rho_{{\vec q},s}^{a}$ of the tagged molecule. 
Normalized tagged-molecule's correlators shall be written as
$\phi_{q,s}^{\rm X}(t) = F_{q,s}^{\rm X}(t)/w_{q}^{\rm X}$
with $w_{q}^{\rm X} = F_{q,s}^{\rm X}(0)$. 

Due to the $C_{2v}$ symmetry of the LW OTP molecule, one finds that
density correlators involving $\rho_{\vec q}^{\rm Q}$, 
$\langle \rho_{\vec q}^{\rm X}(t)^{*} \rho_{\vec q}^{\rm Q} \rangle/N$, 
become nonzero only for ${\rm X} = {\rm Q}$, and the nonzero correlator
for ${\rm X} = {\rm Q}$ is identical to its self-part,
i.e., $F_{q}^{\rm Q}(t) = F_{q,s}^{\rm Q}(t)$.
Viewed as a matrix, this means that
$\langle \rho_{\vec q}^{\rm X}(t)^{*} \rho_{\vec q}^{\rm Y} \rangle/N$
(${\rm X}, {\rm Y} = {\rm N}$, ${\rm Z}$, or ${\rm Q}$)
can be represented as
\begin{equation}
\left(
\begin{array}{ccc}
F_{q}^{\rm N}(t)  & F_{q}^{\rm NZ}(t) & 0 \\
F_{q}^{\rm NZ}(t) & F_{q}^{\rm Z}(t)  & 0 \\
0                 & 0                 & F_{q,s}^{\rm Q}(t)
\end{array}
\right) \, ,
\label{eq:new-representation}
\end{equation}
where the only cross correlation is given by
$F_{q}^{\rm NZ}(t) = 
\langle \rho_{\vec q}^{\rm N}(t)^{*} \rho_{\vec q}^{\rm Z} \rangle/N$.
Thus, the dynamics associated with the variable ${\rm Q}$ 
is uncoupled from the one with ${\rm N}$ and ${\rm Z}$, and is
strictly connected to the single-molecule dynamics. 

Assuming the equal scattering lengths for all the constituent sites, 
the normalized correlator
$\phi_{q}^{\rm N}(t)$ is directly related
to the cross section as measured in the coherent neutron scattering.
The functional forms of $\phi_{q}^{\rm Z}(t)$ and $\phi_{q}^{\rm Q}(t)$
have been chosen so that their
small-wavenumber limits reduce to the 1st-rank 
reorientational correlators~\cite{Chong03}
respectively for the directions associated with
the Y and X axes in Fig.~\ref{fig:picture-OTP}. 
Dynamical features of $\phi_{q}^{\rm N}(t)$ based on MD simulations
have already been discussed in Refs.~\cite{Rinaldi01,Chong03}, and
some results on $\phi_{q}^{\rm Z}(t)$ and $\phi_{q}^{\rm Q}(t)$
in Ref.~\cite{Chong03}. 

Two formulae shall be quoted here which are useful
in analyzing simulation results for correlators. 
(For simplicity, we shall write down only those formulas for 
$\phi_{q}^{\rm X}(t)$, but similar ones hold 
for all the correlators introduced above.)
The correlator $\phi_{q}^{\rm X}(t)$ in supercooled states
exhibits the two-step relaxation:
the relaxation toward the plateau, followed by the
final relaxation from the plateau to zero.
The latter is referred to as the $\alpha$-relaxation.
The von Schweidler law
as derived from MCT
including the next to leading order corrections~\cite{Franosch97}
\begin{equation}
\phi_{q}^{\rm X}(t) = f_{q}^{{\rm X}c} -
h_{q}^{{\rm X}} \, (t/\tau)^{b} +
h_{q}^{{\rm X} (2)} \, (t/\tau)^{2b} +
O((t/\tau)^{3b}),
\label{eq:vs}
\end{equation}
describes
the departure from the plateau value $f_{q}^{{\rm X}c}$
-- also referred to as the critical nonergodicity parameter --
in the early $\alpha$-relaxation region.
Here $h_{q}^{{\rm X}}$ and $h_{q}^{{\rm X} (2)}$ denote
the critical and correction amplitudes, respectively,
$b$ the von Schweidler exponent, and
$\tau$ the $\alpha$-relaxation time.
The exponent $b$ is uniquely related to the so-called
exponent parameter $\lambda$ defined in MCT~\cite{Goetze91b}. 
Another formula,
which is purely empirical and 
well adopted for fitting correlators
in the $\alpha$-relaxation region, is the 
Kohlrausch law
\begin{equation}
\phi_{q}^{\rm X}(t) = A_{q}^{\rm X} \, 
\exp [ - ( t/\tau_{q}^{\rm X} )^{\beta_{q}^{\rm X}} ] \, ,
\label{eq:Kohlrausch}
\end{equation}
with the correlator-dependent 
$\alpha$-relaxation time $\tau_{q}^{\rm X}$ and 
the stretching exponent $\beta_{q}^{\rm X}$. 
It is worthwhile to mention that, in the large-$q$ limit,
the Kohlrausch law (\ref{eq:Kohlrausch}) can be derived
from MCT~\cite{Fuchs94}.
In particular, one gets 
$\lim_{q \to \infty} \beta_{q}^{\rm X} = b$ irrespective of the
choice for ${\rm X}$. 
Since the exponent $b$ is uniquely related to $\lambda$,
the Kohlrausch-law fit in the large-$q$ regime
thus provides a means to estimate $\lambda$
based on simulation results for $\phi_{q}^{\rm X}(t)$. 

\section{Summary of simulation results}
\label{sec:MD}

In this section we summarize results of MD
simulations performed for LW OTP:
more complete description of the simulation results 
can be found in Refs.~\cite{Rinaldi01,Mossa02,Chong03}. 
We also present some additional results which have not been
considered so far.
At the final subsection, we discuss that the simulation 
results for LW OTP share many features with 
experimental data for real system. 

\subsection{Static structure factors}
\label{sec:MD-1}

We first briefly review the main features of the static structure factors, 
which turn out to be important in understanding simulation results for dynamics. 
These static structure factors are also to be used as input in theoretical
calculations presented in Sec.~\ref{sec:MCT}.

\begin{figure}
\includegraphics[width=0.7\linewidth]{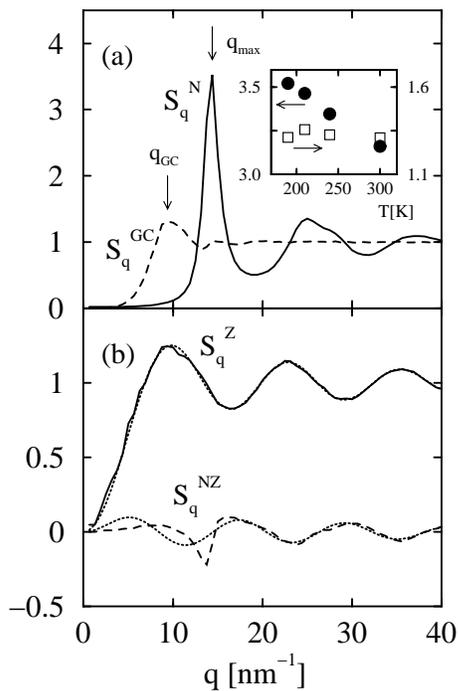}
\caption{
MD simulation results for the static structure factors
at $\rho = 2.71$ molecules$/$nm$^{3}$ and $T = 190$ K.
(a) Solid and dashed lines respectively denote the 
static structure factor $S_{q}^{\rm N}$ and 
the geometrical-center static structure
factor $S_{q}^{\rm GC}$.
In this and the following figures, the arrows indicate the peak positions 
$q_{\rm max}$ ($\approx 14.5$ nm$^{-1}$) in $S_{q}^{\rm N}$ and
$q_{\rm GC}$ ($\approx 9$ nm$^{-1}$) in $S_{q}^{\rm GC}$.
The inset exhibits the temperature dependence of the peak heights 
for $S_{q}^{\rm N}$ (filled circles, left scale) and
$S_{q}^{\rm GC}$ (open squares, right scale). 
(b) Solid and dashed lines respectively denote the static structure
factors $S_{q}^{\rm Z}$ and $S_{q}^{\rm NZ}$. 
Dotted lines refer to their self-parts, $w_{q}^{\rm Z}$ and $w_{q}^{\rm NZ}$.}
\label{fig:static-MD}
\end{figure}

Representative MD simulation results for the static structure factors 
in a supercooled state are presented in 
Figs.~\ref{fig:static-MD}(a) and (b).
We show in Fig.~\ref{fig:static-MD}(a)
the static structure factor $S_{q}^{\rm N}$ and the
geometrical-center (GC) static structure factor $S_{q}^{\rm GC}$.
The latter is defined by
$S_{q}^{\rm GC} =
\sum_{j,l}
\langle 
e^{- i {\vec q} \cdot ({\vec r}_{j}^{\, \rm GC} - {\vec r}_{l}^{\, \rm GC})}
\rangle /N$
with ${\vec r}_{j}^{\, \rm GC}$ denoting the
GC position ${\vec r}_{j}^{\, \rm GC} = (1/3) \sum_{a=1}^{3} {\vec r}_{j}^{\, a}$
of the LW OTP molecule. 
The GC position actually coincides with the center-of-mass position,
but we prefer to use the term ``GC'' because of the reason 
discussed in Ref.~\cite{Chong03}. 
$S_{q}^{\rm N}$ has a main peak at $q = q_{\rm max}$ ($\approx 14.5$ nm$^{-1}$),
which is compatible with the inverse of the LJ diameter 
$\sigma = 0.483$ nm of a site in the sense that 
$q_{\rm max} \sigma \approx 7$. 
On the other hand,
$S_{q}^{\rm GC}$ has a peak at $q = q_{\rm GC}$ ($\approx 9$ nm$^{-1}$)
which can be related to the inverse of the van der Waals radius
$r_{\rm W} = 0.37$ nm for OTP molecule~\cite{Bondi64} 
since $q_{\rm GC} \times (2 r_{\rm W}) \approx 7$,
i.e., it is connected to the overall size of the molecule
({\em cf}. Fig.~\ref{fig:picture-OTP}). 

Figure~\ref{fig:static-MD}(b) exhibits the static structure factor
$S_{q}^{\rm Z}$ and the cross correlation $S_{q}^{\rm NZ} = F_{q}^{\rm NZ}(0)$.
Also added in this figure are their self-parts,
$w_{q}^{\rm Z}$ and $w_{q}^{\rm NZ}$.
One recognizes that $S_{q}^{\rm Z}$ and $w_{q}^{\rm Z}$ are
nearly the same.
It is also seen that $S_{q}^{\rm NZ} \approx w_{q}^{NZ}$
holds to a reasonable extent, and that the magnitude of the
cross correlation is rather small. 
This means that, within the description based on the site-site
static structure factors, 
the representation~(\ref{eq:new-representation}) for $t = 0$ is 
nearly diagonal, 
in which essentially only $S_{q}^{\rm N}$ is associated with
the intermolecular correlation. 

The peak at $q = q_{\rm GC}$ in $S_{q}^{\rm GC}$ also reflects 
intermolecular static correlations in the system. 
However, as discussed in Ref.~\cite{Rinaldi01} and reproduced
in the inset of Fig.~\ref{fig:static-MD}(a), 
the most pronounced
temperature dependence in the static structure factors shows up
at $q = q_{\rm max}$ in $S_{q}^{\rm N}$, 
whereas the peak at $q = q_{\rm GC}$ in $S_{q}^{\rm GC}$ is 
nearly temperature independent.
Thus, for LW OTP, 
the slowing down of the dynamics upon lowering $T$ is 
connected with the intermolecular correlation 
manifested as the main peak in $S_{q}^{\rm N}$ (the cage effect). 

In the present work, we shall primarily be interested in 
the collective dynamics arising from intermolecular correlations. 
As noticed in connection with Eq.~(\ref{eq:new-representation}),
the correlator $\phi_{q}^{\rm Q}(t)$ is strictly 
connected to the single-molecule dynamics, 
$\phi_{q}^{\rm Q}(t) = \phi_{q,s}^{\rm Q}(t)$.
Furthermore, $S_{q}^{\rm Z} \approx w_{q}^{\rm Z}$ and 
$S_{q}^{\rm NZ} \approx w_{q}^{\rm NZ} \approx 0$ 
shown in Fig.~\ref{fig:static-MD}(b)
imply that also the correlator $\phi_{q}^{\rm Z}(t)$ 
approximately reflects the single-molecule dynamics only,  
i.e., $\phi_{q}^{\rm Z}(t) \approx \phi_{q,s}^{\rm Z}(t)$. 
Indeed, we confirmed both from simulation and theoretical 
results that such approximation holds to a very good extent
for the whole time region.
We shall therefore not consider the correlators 
$\phi_{q}^{\rm Z}(t)$ and $\phi_{q}^{\rm Q}(t)$ any more 
in the following. 

\subsection{Density dependence of 
$\mbox{\boldmath $T_{c}$}$ and 
$\mbox{\boldmath $\lambda$}$}
\label{sec:MD-2}

\begin{figure}
\includegraphics[width=0.7\linewidth]{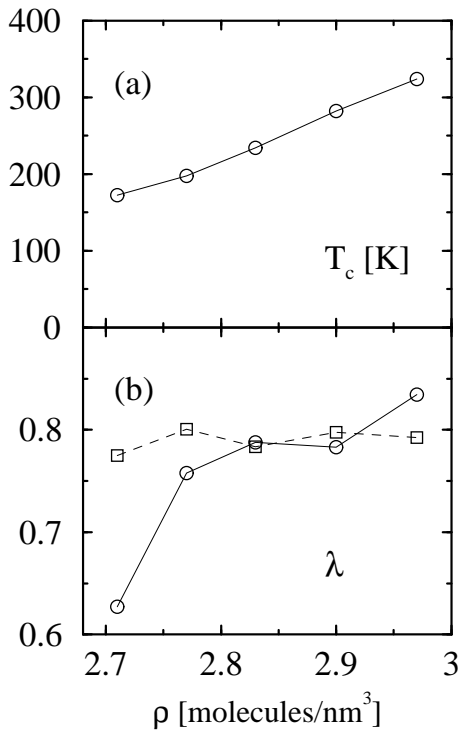}
\caption{
MD simulation results for (a) the MCT critical temperature $T_{c}$
and (b) the exponent parameter $\lambda$ as a function of 
density $\rho$.
Circles in (a) and (b) denote $T_{c}$ and $\lambda$
as determined from
the fit of the diffusion constants $D$ according to 
Eq.~(\protect\ref{eq:D-fit}).
In (b), squares refer to $\lambda$ based on the exponent $b$,
which is determined from the large-$q$ behavior of the 
Kohlrausch stretching exponents
$\lim_{q \to \infty} \beta_{q}^{\rm N} = b$
for the density correlators $\phi_{q}^{\rm N}(t)$.
In both (a) and (b), lines are guide to the eyes.}
\label{fig:Tc-lambda-MD}
\end{figure}

Simulation results for the MCT critical temperature $T_{c}$ and the exponent
parameter $\lambda$ are often determined from the fit of the diffusion 
constants $D$ according to the
MCT asymptotic power law~\cite{Goetze91b}
\begin{equation}
D(T) \propto (T - T_{c})^{\gamma},
\label{eq:D-fit}
\end{equation}
where the exponent $\gamma$ is uniquely related to $\lambda$. 
Such a fit has been performed in Ref.~\cite{Mossa02} for various densities, and
circles in Figs.~\ref{fig:Tc-lambda-MD}(a) and (b) 
denote the resulting $T_{c}$ and $\lambda$ as a function
of the density $\rho$. 
The critical temperature $T_{c}$ increases with 
increasing $\rho$. 
The exponent parameter $\lambda$ seems to increase with 
increasing $\rho$, but the statistical errors in simulations
do not allow to rule out the possibility of
a constant value (see below).
We also notice that an unbiased three-parameter 
fit based on Eq.~(\ref{eq:D-fit}) 
suffers from correlations between the fit parameters 
$T_{c}$ and $\gamma$~\cite{Foffi03}. 
It also suffers from the identification of a fitting $T$ range 
bounded below from $T_{c}$ and above from the $T$ at which 
correlation functions start to obey the time-temperature superposition.
We note in passing that 
deviation from the MCT behavior for $T<T_c$ has been 
discussed for the LW-OTP model in Ref.~\cite{Mossa02}.

Often, the $\rho$ dependence of $T_{c}$ 
can be condensed into an effective coupling parameter
\begin{equation}
\Gamma \propto \rho T^{-1/4} \, .
\label{eq:effective-coupling}
\end{equation}
This is the only relevant parameter for specifying the thermodynamic state
of soft-sphere systems whose 
repulsive interaction is proportional to $r^{-12}$~\cite{Hansen86}.
Indeed, in the case of binary mixture of soft spheres,
it was found from computer simulations that the ideal glass transition
occurs at a constant value, $\Gamma = \Gamma_{c}$~\cite{Bernu87,Roux89}.
Although Eq.~(\ref{eq:effective-coupling}) is valid only for the
$r^{-12}$ soft-sphere system, 
a computer-simulation study for 
a model of polymer at different pressures~\cite{Bennemann99c}
indicate that, also for LJ systems, the MCT critical point 
might be described well in terms of the effective coupling 
parameter. 
For LW OTP, we find from the $\rho$ dependence of $T_{c}$ shown in 
Fig.~\ref{fig:Tc-lambda-MD}(a) that
$\Gamma_{c} = 1.57 \pm 0.05$, where the
prefactor in Eq.~(\ref{eq:effective-coupling}) has been chosen 
to be $\sigma_{\Gamma}^{3} \epsilon_{\Gamma}^{1/4}$ 
with $\sigma_{\Gamma} = 0.76$ nm and $\epsilon_{\Gamma} = 600$ K
as in Ref.~\cite{Toelle98b} for later comparison with experimental result. 
Thus, to a reasonable extent, the parameter $\Gamma$ for LW OTP is also
found to be nearly constant at the MCT critical point. 

Concerning the $\rho$ dependence of the
exponent parameter $\lambda$ shown in Fig.~\ref{fig:Tc-lambda-MD}(b), 
we notice that
a different value for $\lambda$ is occasionally obtained 
from some other analysis of simulation results.
For example, one gets another estimate for
$\lambda$ from the Kohlrausch-law fit of some density correlators 
as described just after Eq.~(\ref{eq:Kohlrausch}).
$\lambda$ obtained in this way, based on the Kohlrausch-law
fit of the correlators $\phi_{q}^{\rm N}(t)$, 
are also included as squares in Fig.~\ref{fig:Tc-lambda-MD}(b),
which are nearly $\rho$-independent.
(We confirmed that the stretching exponents 
in the large-$q$ regime do not 
depend on the choice of correlators.)
The difference between circles and squares in 
Fig.~\ref{fig:Tc-lambda-MD}(b) can be considered as a sort of error bars
in determining $\lambda$ based on the simulation results,
but we already notice here that the insensitivity of $\lambda$
to density is consistent with the
experimental and theoretical results to be described below. 

\subsection{Critical nonergodicity parameters and critical amplitudes}
\label{sec:MD-3}

\begin{figure}
\includegraphics[width=0.65\linewidth]{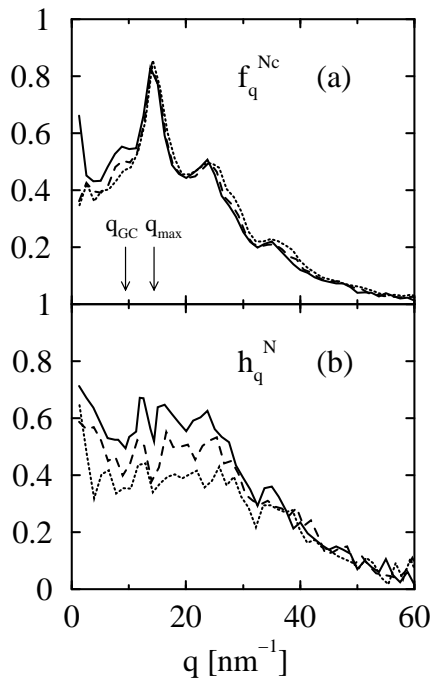}
\caption{
MD simulation results for
(a) the critical nonergodicity parameters 
$f_{q}^{{\rm N}c}$ and
(b) the critical amplitudes $h_{q}^{\rm N}$ 
of the correlators $\phi_{q}^{\rm N}(t)$
as determined from the fit according to 
Eq.~(\protect\ref{eq:vs})
for three densities
$\rho = 2.71$ (solid lines), 2.83 (dashed lines), and
2.97 molecules$/$nm$^{3}$ (dotted lines).
$h_{q}^{\rm N}$ are in arbitrary units.}
\label{fig:fqc-NN-MD}
\end{figure}

Figures~\ref{fig:fqc-NN-MD}(a) and (b) show the $q$ dependence of 
the critical nonergodicity parameters $f_{q}^{{\rm N}c}$ and 
the critical amplitudes $h_{q}^{\rm N}$ of 
the correlators $\phi_{q}^{\rm N}(t)$
based on the fit according to Eq.~(\ref{eq:vs})
for three representative densities,
$\rho = 2.71$, 2.83, and 2.97 molecules$/$nm$^{3}$.
The fit of $\phi_{q}^{\rm N}(t)$
according to Eq.~(\ref{eq:vs}) has been performed
by constraining the exponent $b$ to the value specified by $\lambda$, 
the latter being taken from the squares in Fig.~\ref{fig:Tc-lambda-MD}(b),
and by regarding $f_{q}^{{\rm N}c}$, 
$h_{q}^{\rm N}/\tau^{b}$ and
$h_{q}^{{\rm N}(2)}/\tau^{2b}$ as fitting parameters. 
Thus, $h_{q}^{\rm N}$ and $h_{q}^{{\rm N}(2)}$ can be determined
only up to $q$-independent multiplicative factors, 
and this is why $h_{q}^{\rm N}$ shown in Fig.~\ref{fig:fqc-NN-MD}(b)
are given in arbitrary units. 
(Hence, one cannot directly compare the amplitude but only the wavevector 
dependence in $h_{q}^{\rm N}$ for different densities.)

For $q \gtrsim q_{\rm max}$ 
($\approx 14.5$ nm$^{-1}$ for all the densities
considered),
$f_{q}^{{\rm N}c}$
for the three densities are very similar to each other.
For $q < q_{\rm max}$, on the other hand, 
$f_{q}^{{\rm N}c}$ is higher for lower density, 
and this holds in particular around the intermediate wavenumber 
$q_{\rm GC}$ ($\approx 9$ nm$^{-1}$):
whereas $f_{q}^{{\rm N}c}$ at $q \approx q_{\rm GC}$
exhibits only a shoulder for the highest density 
$\rho = 2.97$ molecules$/$nm$^{3}$,
a well-developed peak is discernible 
for the lowest density $\rho = 2.71$ molecules$/$nm$^{3}$.

The results for $h_{q}^{\rm N}$ shown in Fig.~\ref{fig:fqc-NN-MD}(b) are somewhat noisy compared to $f_{q}^{{\rm N}c}$.
Concerning the $q$ dependence, we notice that
there are two minima in $h_{q}^{\rm N}$ located 
at $q \approx q_{\rm GC}$ and $q_{\rm max}$.
As will be mentioned in Sec.~\ref{sec:MD-4},
this is consistent with the existence of the 
peaks in $f_{q}^{{\rm N}c}$ around these two wavenumbers.
Furthermore, the minimum in $h_{q}^{\rm N}$ at $q \approx q_{\rm GC}$
is more pronounced
for lower density in the sense that 
$h_{q_{\rm max}}^{\rm N}/h_{q_{\rm GC}}^{\rm N}$ is larger
for lower $\rho$, 
and this is also consistent with the found density dependence
of $f_{q}^{{\rm N}c}$ around this wavenumber. 

\subsection{Unusual feature in the wavenumber dependence}
\label{sec:MD-4}

All parameters describing density correlations exhibit
a characteristic wavenumber dependence reflecting that of 
the underlying static structure factor. 
As found in Ref.~\cite{Rinaldi01} and also discussed in more detail in Ref.~\cite{Chong03}, however,
the dynamics in LW OTP exhibits an
unusual wavenumber dependence of the critical
nonergodicity parameters and the $\alpha$-relaxation times
of the correlators $\phi_{q}^{\rm N}(t)$ 
at intermediate wavenumbers $q \approx q_{\rm GC}$. 
Before embarking on the unusual feature, let us first consider 
a ``usual'' case as a reference. 

\begin{figure}
\includegraphics[width=0.7\linewidth]{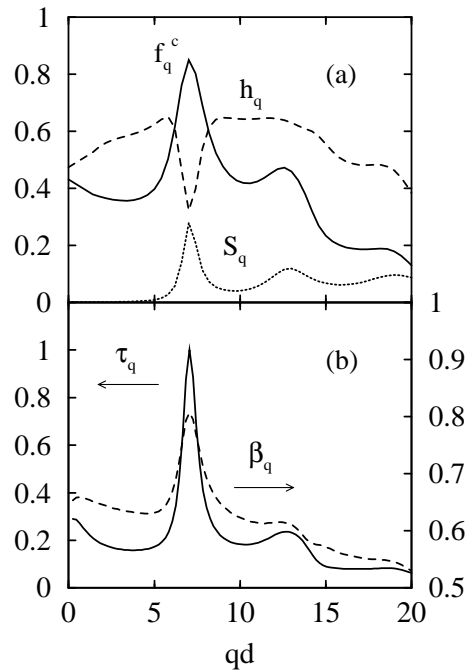}
\caption{
MCT results for the wavenumber dependence of various quantities for the
hard-sphere system as a function of $qd$ with $d$ denoting
the hard-sphere diameter. 
(a) Solid and dashed lines respectively denote 
the critical nonergodicity parameters $f_{q}^{c}$
and the critical amplitudes $h_{q}$ for normalized density
correlators $\phi_{q}(t)$. 
Dotted line refers to the Percus-Yevick static structure 
factor $S_{q}$, multiplied by a factor of 0.08 for ease
of comparison, at the critical point. 
(b) Solid and dashed lines respectively denote the 
$\alpha$-relaxation times $\tau_{q}$ and
the stretching exponents $\beta_{q}$
obtained from the fit according to Eq.~(\protect\ref{eq:Kohlrausch})
of the $\alpha$-master curve for $\phi_{q}(t)$.
$\tau_{q}$ are in arbitrary units.}
\label{fig:fqc-tau-Sq-HSS}
\end{figure}

We show in Figs.~\ref{fig:fqc-tau-Sq-HSS}(a) and (b)
the critical nonergodicity parameters $f_{q}^{c}$,
the critical amplitudes $h_{q}$,
the $\alpha$-relaxation times $\tau_{q}$, and the
Kohlrausch stretching exponents $\beta_{q}$ 
of normalized density correlators $\phi_{q}(t)$
for the hard-sphere system as determined 
by solving the MCT equations for simple systems~\cite{Goetze91b,Fuchs92b}. 
$\tau_{q}$ and $\beta_{q}$ have been obtained from the
fit according to Eq.~(\ref{eq:Kohlrausch}) 
of the $\alpha$-master curve for $\phi_{q}(t)$. 
The static structure factor used is evaluated within the 
Percus-Yevick approximation~\cite{Hansen86},
and the one at the critical point is included in 
Fig.~\ref{fig:fqc-tau-Sq-HSS}(a). 
A strong correlation in the $q$ dependence of these quantities 
can clearly be observed for the whole wavenumber regime: 
$f_{q}^{c}$, $1/h_{q}$, $\tau_{q}$, and $\beta_{q}$ oscillate 
in phase with $S_{q}$, and this holds in particular around 
the first sharp diffraction peak, $qd \approx 7$ with $d$ 
denoting the hard-sphere diameter.
A theoretical explanation of such correlation has already
been documented~\cite{Franosch97,Goetze00c},
and shall not be repeated here. 
Such ``usual'' results have been observed in 
simulation studies for LJ binary mixture~\cite{Kob95b}, 
silica~\cite{Horbach01}, and water~\cite{Sciortino97}, and also 
in experimental results reviewed in Ref.~\cite{Petry95}. 

\begin{figure}
\includegraphics[width=0.7\linewidth]{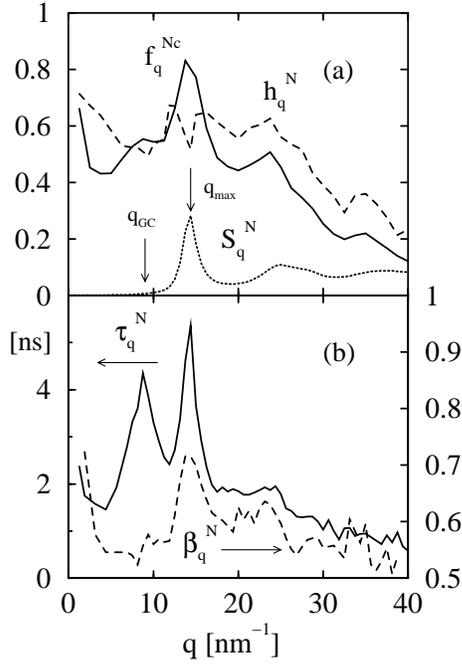}
\caption{
MD simulation results for the wavenumber dependence 
of various quantities for LW OTP at 
$\rho = 2.71$ molecules$/$nm$^{3}$. 
(a) Solid and dashed lines respectively denote 
the critical nonergodicity parameters $f_{q}^{{\rm N}c}$
and the critical amplitudes $h_{q}^{\rm N}$ for the
correlators $\phi_{q}^{\rm N}(t)$. 
$h_{q}^{\rm N}$ are in arbitrary units. 
Dotted line refers to the static structure 
factor $S_{q}^{\rm N}$, multiplied by a factor of 0.08 for ease
of comparison, at $T = 190$~K.
({\em cf.} $T_{c} \approx 172$~K for this density, 
see Fig.~\protect\ref{fig:Tc-lambda-MD}.)
(b) Solid and dashed lines respectively denote 
the $\alpha$-relaxation times $\tau_{q}$ and
the stretching exponents $\beta_{q}$ 
obtained from the fit according to Eq.~(\protect\ref{eq:Kohlrausch})
of the correlators $\phi_{q}^{\rm N}(t)$ at $T=190$~K.}
\label{fig:fqc-tau-Sq-271-MD}
\end{figure}

The result found in the simulation study for LW OTP 
is unusual in that such a correlation is violated 
at intermediate wavenumbers $q \approx q_{\rm GC}$. 
This is summarized in Figs.~\ref{fig:fqc-tau-Sq-271-MD}(a) and (b) for the density $\rho = 2.71$ molecules$/$nm$^{3}$. 
A related figure for $\rho = 2.83$ molecules$/$nm$^{3}$
can be found in Ref.~\cite{Chong03}. 
One recognizes from Figs.~\ref{fig:fqc-tau-Sq-271-MD}(a) and (b) that $f_{q}^{{\rm N}c}$, $1/h_{q}^{\rm N}$ and $\tau_{q}^{\rm N}$ of the
correlator $\phi_{q}^{\rm N}(t)$ exhibit an additional peak
at $q \approx q_{\rm GC}$ which does not exist in the
corresponding static structure factor $S_{q}^{\rm N}$.
A signature of such a peak in $\beta_{q}^{\rm N}$ is also 
discernible, but a definitive conclusion cannot be drawn 
since the results for $\beta_{q}^{\rm N}$ are rather noisy. 
As already noticed in connection with 
Figs.~\ref{fig:fqc-NN-MD}(a) and (b), this unusual feature is 
found to be more pronounced for lower density.
It was suggested in Refs.~\cite{Rinaldi01,Chong03} that 
the unusual feature is caused by the coupling of 
the rotational motion to the GC motion.
This follows from the fact that the GC static structure factor
$S_{q}^{\rm GC}$ has a peak at $q = q_{\rm GC}$
as shown in Fig.~\ref{fig:static-MD}(a).

\begin{figure}
\includegraphics[width=0.6\linewidth]{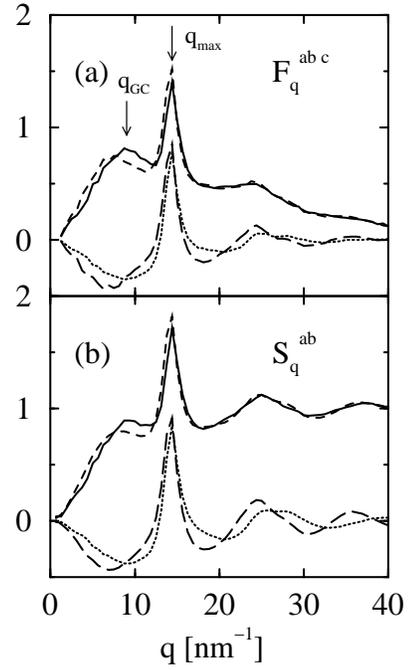}
\caption{
MD simulation results for 
the site-site critical nonergodicity parameters
$F_{q}^{ab \, c}$
and the site-site static structure factors
$S_{q}^{ab}$ for the density $\rho = 2.71$ molecules$/$nm$^{3}$. 
(a) 
$F_{q}^{11 \, c}$ (solid line), 
$F_{q}^{12 \, c}$ (dotted line), 
$F_{q}^{22 \, c}$ (dashed line), and
$F_{q}^{23 \, c}$ (long-dashed line).
(b) 
$S_{q}^{11}$ (solid line), 
$S_{q}^{12}$ (dotted line), 
$S_{q}^{22}$ (dashed line), and
$S_{q}^{23}$ (long-dashed line)
at $T = 190$ K.}
\label{fig:fqc-Fqc-Sq-site-MD}
\end{figure}

To understand what is really the unusual feature, 
especially concerning $f_{q}^{{\rm N}c}$ shown in 
Fig.~\ref{fig:fqc-tau-Sq-271-MD}(a),
we go back to the original site-site density correlators $F_{q}^{ab}(t)$. 
Let us recall that 
$\phi_{q}^{\rm N}(t) = F_{q}^{\rm N}(t) / S_{q}^{\rm N}$ with
$F_{q}^{\rm N}(t) = \sum_{a,b=1}^{3} F_{q}^{ab}(t)/3$ and
$S_{q}^{\rm N} = \sum_{a,b=1}^{3} S_{q}^{ab}/3$.
Critical nonergodicity parameters $F_{q}^{ab \, c}$ of the
correlators $F_{q}^{ab}(t)$, obtained similarly from the fit according to
Eq.~(\ref{eq:vs}), and the site-site static structure factors
$S_{q}^{ab}$ are shown in 
Figs.~\ref{fig:fqc-Fqc-Sq-site-MD}(a) and (b),
respectively.
(Only the independent components in $F_{q}^{ab \, c}$ and $S_{q}^{ab}$
are shown.
The independent components, e.g., in the total nine $S_{q}^{ab}$, are
$S_{q}^{11}$, $S_{q}^{12}$, $S_{q}^{22}$, and $S_{q}^{23}$ 
due to the symmetry $S_{q}^{ab} = S_{q}^{ba}$ as well
as the $C_{2v}$ symmetry of the LW OTP molecule.)
It is seen from the comparison of 
Figs.~\ref{fig:fqc-Fqc-Sq-site-MD}(a) and (b) that
the results for LW OTP are ``usual'' in the sense that
the $q$ dependence of $F_{q}^{ab \, c}$ correlates well with
that of $S_{q}^{ab}$ in the whole wavenumber regime
including $q \approx q_{\rm GC}$. 
We also notice that these quantities take 
positive and negative
values at $q \approx q_{\rm GC}$. 
The unusual feature concerning $f_{q}^{{\rm N}c}$
shows up only after summing
up the site-site $F_{q}^{ab \, c}$ and $S_{q}^{ab}$
to obtain $F_{q}^{{\rm N}c}$ and $S_{q}^{\rm N}$. 
One observes that positive and negative components in 
$S_{q}^{ab}$ almost cancel out after taking the summation, 
which results in small and flat $S_{q}^{\rm N}$ at 
$q \approx q_{\rm GC}$ as shown in 
Fig.~\ref{fig:fqc-tau-Sq-271-MD}(a). 
On the other hand, such a cancellation does not occur
in the summation of the components $F_{q}^{ab \, c}$, 
and this causes the unusual peak 
in $F_{q}^{{\rm N}c}$, and hence in 
$f_{q}^{{\rm N}c} = F_{q}^{{\rm N}c}/S_{q}^{\rm N}$,
at $q \approx q_{\rm GC}$. 
Thus, the mentioned unusual feature reflects purely dynamical effects, 
since it can be observed only in such dynamical quantities
as $f_{q}^{{\rm N}c}$, $1/h_{q}^{\rm N}$ and $\tau_{q}^{\rm N}$. 

\subsection{Comparison with experimental results}
\label{sec:MD-5}

Sine the LW OTP model is a very simplified one, 
one might think that a straightforward comparison of 
the MD simulation results with experimental data is not feasible. 
However, we discuss in the following that 
the simulation results for LW OTP share many features with 
experimental ones. 
Since we are interested in the $q$ dependence of 
collective dynamical quantities, especially in the vicinity
of $q_{\rm GC}$ and $q_{\rm max}$,
we shall be mostly concerned
with coherent neutron-scattering results. 
A review of neutron-scattering studies of OTP is presented
in Ref.~\cite{Toelle01}, and the experimental results
to be presented below can be found in this article and 
references cited therein. 

Let us first consider the static structure factor 
$S_{q}^{\rm exp}$ from coherent neutron scattering for 
fully deuterated OTP, which is given as a weighted sum of atomic correlations, 
$S_{q}^{\rm exp} \propto \sum_{\alpha, \beta} 
b_{\alpha} b_{\beta} S_{q}^{\alpha \beta}$.
Here $\alpha$ and $\beta$ refer to deuteron and carbon atoms,
$b_{\alpha}$ the scattering length,
and $S_{q}^{\alpha \beta}$ the partial structure factors.
(Greek characters $\alpha$ and $\beta$ are used here to distinguish
them from Roman characters $a$ and $b$ 
which have been adopted to label sites in the LW OTP molecule.
Also, quantities as determined from experiments
shall be distinguished with the superscript ``exp''.)
It has been observed that, in contrast to atomic systems, the main peak of
$S_{q}^{\rm exp}$ is split into two maxima at about 
$q = 14$ and 19 nm$^{-1}$~\cite{Toelle97}. 
Also, a shallow shoulder is discernible in $S_{q}^{\rm exp}$ around 
$q = 9$ nm$^{-1}$. 
It was conjectured that the peak at $q \approx 19$ nm$^{-1}$ is
built up mainly by intramolecular correlations within phenyl rings,
while the maximum at $q \approx 14$ nm$^{-1}$ is associated with
intermolecular correlations between phenyl rings.
The shoulder at $q \approx 9$ nm$^{-1}$ was interpreted 
as being due to correlations between molecular centers of mass 
since its position is compatible with the inverse of the van der
Waals radius $r_{\rm W} = 0.37$ nm~\cite{Bondi64}.

This picture for $S_{q}^{\rm exp}$ is consistent with the
simulation results for $S_{q}^{\rm N}$ and $S_{q}^{\rm GC}$ 
shown in Fig.~\ref{fig:static-MD}(a). 
Since each site of the LW OTP molecule represents an entire
phenyl ring, the peak at $q_{\rm max} \approx 14.5$ nm$^{-1}$
in $S_{q}^{\rm N}$ corresponds to the first main peak in $S_{q}^{\rm exp}$.
No second peak around $q \approx 19$ nm$^{-1}$ can be observed in
$S_{q}^{\rm N}$ since the internal structure within a phenyl ring 
is completely discarded in the LW OTP model.
The peak at $q_{\rm GC} \approx 9$ nm$^{-1}$ in $S_{q}^{\rm GC}$ is 
related to the shoulder in $S_{q}^{\rm exp}$, 
although this is not reflected in $S_{q}^{\rm N}$. 
As discussed in connection with Fig.~\ref{fig:fqc-Fqc-Sq-site-MD}(b), 
the disappearance of any shoulder or peak at $q \approx q_{\rm GC}$
in $S_{q}^{\rm N}$ is due to the cancellation of the constituent
site-site correlation functions $S_{q}^{ab}$, 
which do exhibit (positive and negative) peaks around this wavenumber. 
Such an almost perfect cancellation might not occur in $S_{q}^{\rm exp}$,
due to more complicated nature of the constituent atomic correlations
$S_{q}^{\alpha \beta}$ in real system. 
Indeed, this conjecture is supported by MD-simulation
studies for more realistic OTP models~\cite{Kudchadkar95,Mossa-OTP-all},
where a shoulder at $q \approx 9$ nm$^{-1}$ is discernible
in the static structure factor which corresponds to $S_{q}^{\rm exp}$.

It has been discussed for a liquid of linear molecules
that orientational correlations can lead to a prepeak at low $q$~\cite{Theis99}.
By calculating the corresponding static orientational correlation functions, 
we confirmed that there is no prominent peak at $q_{\rm GC} \approx 9$ nm$^{-1}$
in these functions for LW OTP. 
Therefore, we do not think the shoulder at $q \approx 9$ nm$^{-1}$
as observed in $S_{q}^{\rm exp}$ reflects the orientational correlations
of the kind discussed in Ref.~\cite{Theis99},
and this is consistent with the picture that the shoulder
stems from the correlations between the molecular centers of mass. 

We next compare the simulation and experimental results 
for the $\rho$ dependence of $T_{c}$ and $\lambda$.
It has been shown in Ref.~\cite{Toelle98b} from an analysis of 
incoherent density correlators at various pressures that
the $\rho$ dependence of $T_{c}$ can be combined to the
effective coupling parameter given in Eq.~(\ref{eq:effective-coupling}). 
This observation is in agreement with the simulation result for LW OTP.
Furthermore, using the same prefactor as described just after 
Eq.~(\ref{eq:effective-coupling}),
the experimental value 
$\Gamma_{c}^{\rm exp} \approx 1.498 \pm 0.004$ 
characterizing the MCT critical point is rather close to
$\Gamma_{c} \approx 1.57 \pm 0.05$ found for LW OTP.
A smaller error bar in $\Gamma_{c}^{\rm exp}$ might be due to a
narrower density range investigated in the experiment. 
Concerning $\lambda$, its insensitivity to density has been
demonstrated with the experimental value $\lambda^{\rm exp} \approx 0.77$. 
This is in accord with the simulation result shown as squares in 
Fig.~\ref{fig:Tc-lambda-MD}(b), including the value for $\lambda$. 

Coherent as well as incoherent neutron-scattering results for
density correlators in supercooled states
exhibit two-step relaxation in agreement with
the prediction of MCT.
The wavenumber dependence of various quantities 
characterizing such glassy dynamics has been 
determined from fits of those density correlators according to 
MCT asymptotic formulas or to the Kohlrausch law,
like we did for simulation results. 
It was observed that the 
critical nonergodicity parameters $f_{q}^{{\rm exp} \, c}$
of coherent density correlators oscillate in phase with
$S_{q}^{\rm exp}$, with the two peaks around $q_{\rm GC}$ and 
$q_{\rm max}$~\cite{Toelle97}.
The critical amplitudes $h_{q}^{\rm exp}$ were found to oscillate
in antiphase with $S_{q}^{\rm exp}$, with the existence of 
two minima at $q \approx q_{\rm GC}$ and $q_{\rm max}$. 
These trends are in agreement with the simulation 
results for $\phi_{q}^{\rm N}(t)$ shown in 
Fig.~\ref{fig:fqc-tau-Sq-271-MD}(a). 

Let us now consider the $q$ dependence of 
the relaxation times and the stretching exponents
of coherent density correlators in the $\alpha$-relaxation regime. 
Some reservation is necessary in the experimental results for 
small $q$ due to the presence of incoherent background 
and to contributions from multiple scattering.
Therefore, the experimental $\alpha$-relaxation times
$\tau_{q}^{\rm exp}$ show a tendency to increase for decreasing $q$. 
Nevertheless, on top of such background, it was observed that
$\tau_{q}^{\rm exp}$ exhibit two plateaus around
$q_{\rm GC}$ and $q_{\rm max}$~\cite{Toelle98}.
This is a signature of the characteristic $q$ dependence 
as found in $\tau_{q}^{\rm N}$ for LW OTP shown in 
Fig.~\ref{fig:fqc-tau-Sq-271-MD}(b).
Concerning the stretching exponents $\beta_{q}^{\rm exp}$, 
a systematic variation in phase with $S_{q}^{\rm exp}$ was observed, 
especially in the vicinity of $q_{\rm max}$~\cite{Toelle98},
although a definitive conclusion cannot be drawn for $q \approx q_{\rm GC}$
due to the noise in the experimental results.
Thus, the overall $q$ dependence of $\tau_{q}^{\rm exp}$ 
and $\beta_{q}^{\rm exp}$ is in accord with the one for LW OTP
shown in Fig.~\ref{fig:fqc-tau-Sq-271-MD}(b),
and even a signature of the unusual peak at $q \approx q_{\rm GC}$
as discussed for LW OTP can be observed in $\tau_{q}^{\rm exp}$.

A natural question arises as to whether the peaks
at $q \approx q_{\rm GC}$ found in the experimental results
for $f_{q}^{{\rm exp} \, c}$, $1/h_{q}^{\rm exp}$ and
$\tau_{q}^{\rm exp}$ can really be considered as unusual. 
This is because, in contrast to $S_{q}^{\rm N}$ for LW OTP,
the experimental static structure factor $S_{q}^{\rm exp}$
exhibits a shoulder, though tiny, in this wavenumber regime.
However, it is rather surprising that such a tiny shoulder in 
$S_{q}^{\rm exp}$ is related to a pronounced wavenumber variation of 
those dynamical quantities, and it seems worthwhile to pay
special attention to dynamical features around $q_{\rm GC}$.
Therefore, we consider that a further investigation for the
unusual features in LW OTP is valuable and might also be
relevant in understanding the experimental results.

\section{Theoretical results}
\label{sec:MCT}

In this section, results for dynamical quantities
calculated from MCT based on the
site representation are presented, and are compared with those 
from the MD simulations. In particular, we examine whether the 
unusual feature at intermediate wavenumbers discussed in Sec.~\ref{sec:MD-4}
can be accounted for by the theory. 
Since the unusual feature is found to be
more pronounced for lower density,
most of the theoretical calculations shall be done for
the lowest density $\rho = 2.71$ molecules/nm$^{3}$
studied in the MD simulations.

\subsection{MCT equations based on the site representation}
\label{sec:MCT-1}

Within the site representation for molecules,
the dynamics of the system is most naturally 
characterized by the site-site density correlators $F_{q}^{ab}(t)$
defined for $a,b = 1, \cdots, n$, where $n$ denotes the number
of sites in a molecule. 
The MCT equations for 
$F_{q}^{ab}(t)$ consist of an exact Zwanzig-Mori equation
and an approximate expression for the relaxation kernel,
which are given in Ref.~\cite{Chong02}.
Regarding $F_{q}^{ab}(t)$ as elements of an 
$n \times n$ matrix ${\bf F}_{q}(t)$, 
the Zwanzig-Mori equation is given by
\begin{subequations}
\label{eq:GLE-v}
\begin{equation}
\partial_{t}^{2} {\bf F}_{q}(t) + {\bf \Omega}_{q}^{2} \, {\bf F}_{q}(t) +
{\bf \Omega}_{q}^{2} 
\int_{0}^{t} dt' \, {\bf m}_{q}(t-t') \, \partial_{t'} {\bf F}_{q}(t') = {\bf 0},
\label{eq:GLE-v-a}
\end{equation}
where ${\bf \Omega}_{q}^{2}$ denotes the characteristic frequency matrix
\begin{equation}
{\bf \Omega}_{q}^{2} = q^{2} \, {\bf J}_{q} \, {\bf S}^{-1}_{q},
\label{eq:GLE-v-b}
\end{equation}
\end{subequations}
with ${\bf S}_{q}^{-1}$ representing the inverse matrix of ${\bf S}_{q}$. 
$J_{q}^{ab}$ denote the site-site static (longitudinal) current
correlation functions, whose explicit expressions for molecules 
possessing the $C_{2v}$ symmetry (like LW OTP and water)
in terms of molecule's inertia parameters can be found in Ref.~\cite{Chong99}.
The MCT expression for the relaxation kernel reads
\begin{subequations}
\label{eq:MCT-v}
\begin{equation}
{\bf m}_{q}(t) = {\bf S}_{q} \, \mbox{\boldmath ${\cal F}$}_{q}[{\bf F}(t)] \, ,
\label{eq:MCT-v-a}
\end{equation}
where the mode-coupling functional 
$\mbox{\boldmath ${\cal F}$}_{q}$ is given by
the equilibrium quantities:
\begin{eqnarray}
& &
\hspace{-0.7cm}
{\cal F}^{ab}_{q}[\, \tilde{\textit {\textbf f}} \, ] =
\frac{1}{2}  
\sum_{\lambda, \mu, \lambda', \mu'}
\int d{\vec k} 
V_{\lambda \mu \lambda' \mu'}^{ab}({\vec q}; {\vec k}, {\vec p} \,) 
\tilde{f}_{k}^{\lambda \mu} 
\tilde{f}_{p}^{\lambda' \mu'},
\label{eq:MCT-v-b}
\\
& &
\hspace{-0.7cm}
V_{\lambda \mu \lambda' \mu'}^{ab}({\vec q}; {\vec k}, {\vec p} \,) =
\frac{\rho}{(2 \pi)^{3}} 
\{ {\vec q} \cdot [\delta^{a \lambda'} {\vec k} \, c_{k}^{a \lambda} +
                   \delta^{a \lambda}  {\vec p} \, c_{p}^{a \lambda'}  ] \} 
\nonumber \\
& &
\qquad \qquad 
\times \,
\{ {\vec q} \cdot [\delta^{b \mu'}     {\vec k} \, c_{k}^{b \mu} +
                   \delta^{b \mu}      {\vec p} \, c_{p}^{b \mu'}      ] \} 
/ q^{4},
\label{eq:MCT-v-c}
\end{eqnarray}
\end{subequations}
with ${\vec p} = {\vec q} - {\vec k}$.
(We have used here a slightly different convention for writing 
the mode-coupling functional from the one adopted in Ref.~\cite{Chong02}
to simplify some equations which follow.)
Here, the direct correlation function is defined via 
the site-site
Ornstein-Zernike equation~\cite{Hansen86},
$\rho c_{q}^{ab} = [{\bf w}_{q}^{-1}]^{ab} - [{\bf S}_{q}^{-1}]^{ab}$.
Equations~(\ref{eq:GLE-v}) and (\ref{eq:MCT-v}) constitute a set of
closed equations of motion for determining ${\bf F}_{q}(t)$,
provided the static structure factors $S_{q}^{ab}$ are known.
In the present work, $S_{q}^{ab}$ determined from 
MD simulations shall be used. 

The matrix of long-time limits (or the nonergodicity parameters),
${\bf F}_{q} = {\bf F}_{q}(t \to \infty)$, 
obeys the implicit
equation defined by the mode-coupling functional 
$\mbox{\boldmath ${\cal F}$}_{q}$,
\begin{equation}
{\bf F}_{q} \, [{\bf S}_{q} - {\bf F}_{q}]^{-1} = 
{\bf S}_{q} \, \mbox{\boldmath ${\cal F}$}_{q}[{\bf F}] \, .
\label{eq:DW-v}
\end{equation}
This equation can be derived from 
Eqs.~(\ref{eq:GLE-v-a}) and (\ref{eq:MCT-v-a}) 
by taking the $t \to \infty$ limit. 
From an iterative procedure
${\bf F}_{q}^{(j+1)}[{\bf S}_{q} - {\bf F}_{q}^{(j+1)}]^{-1} =
{\bf S}_{q} \mbox{\boldmath ${\cal F}$}_{q}[{\bf F}^{(j)}]$
starting with ${\bf F}_{q}^{(0)} = {\bf S}_{q}$,
one obtains a solution of Eq.~(\ref{eq:DW-v}) as
${\bf F}_{q} = \lim_{j \to \infty} {\bf F}_{q}^{(j)}$~\cite{Chong02}.
One gets trivial solutions ${\bf F}_{q} = {\bf 0}$ for $T > T_{c}$, whereas
nontrivial solutions ${\bf F}_{q} \succ {\bf 0}$ can be obtained 
for $T \le T_{c}$.
Here, and in the following,
we mean by ${\bf F}_{q} \succ {\bf 0}$ (or $F_{q}^{ab} \succ 0$)
that the matrix ${\bf F}_{q}$ is positive definite.
Thus, one obtains $T_{c}$ 
as the highest temperature at which there holds
${\bf F}_{q} \succ 0$, 
and the solution at this critical point provides the critical 
nonergodicity parameters ${\bf F}_{q}^{c}$.

The convergence of the iterative procedure for Eq.~(\ref{eq:DW-v})
is ruled by the spectral radius of a stability matrix, which can be defined
from the mode-coupling functional $\mbox{\boldmath ${\cal F}$}_{q}$
as in the case of simple systems~\cite{Goetze91b} and is given by
\begin{eqnarray}
C_{qk}^{aba'b'} &=& 
\sum_{p} 
\sum_{\lambda,\mu,\lambda',\mu'}
V_{\lambda \lambda' \mu \mu' \, ; \, qkp}^{ab} 
\nonumber \\
& &
\quad \times \,
[{\bf S}_{k} - {\bf F}_{k}]^{\lambda a'} \, 
[{\bf S}_{k} - {\bf F}_{k}]^{\mu b'} \,
F_{p}^{\lambda' \mu'} \, .
\label{eq:stability}
\end{eqnarray}
In deriving this expression from Eq.~(\ref{eq:MCT-v-b}),
the wavevector integrals are converted into discrete sums
by introducing some upper cutoff and using a grid
of $M$ equally spaced values for the wavenumbers.
Thus, the wavenumber ($q$, $k$ and $p$) can now be
considered as a label for an array of $M$ values. 
Correspondingly, the coefficients 
$V_{\lambda \mu \lambda' \mu'}^{ab}({\vec q}; {\vec k}, {\vec p} \,)$
in Eq.~(\ref{eq:MCT-v-b}) are expressed as
$V_{\lambda \lambda' \mu \mu' \, ; \, qkp}^{ab}$ in Eq.~(\ref{eq:stability}). 
The details of the transformation of the mode-coupling functional
to a polynomial in the discretized variables can be found
in Ref.~\cite{Franosch97}. 

For the notational simplicity, we introduce new indices
$i = (q,a,b)$ and $j = (k,a',b')$ using the so-called dictionary order, which run from 1 to $Mn^{2}$. 
Then, the stability matrix given in Eq.~(\ref{eq:stability})
is simply denoted as $C_{ij}$. 
Unlike the case of simple systems for which the stability matrix is 
given by positive matrix~\cite{Goetze91b}, 
each element of $C_{ij}$ can take positive and negative values. 
However, $C_{ij}$ can be considered as a generalized positive matrix
in the sense that it transforms a positive definite matrix into
another one:
if $x_{i} \succ 0$ (meaning $x_{q}^{ab} \succ 0$), 
then $\sum_{j} C_{ij} x_{j} \succ0$ and
$\sum_{i} x_{i} C_{ij} \succ 0$.
Such matrix $C_{ij}$ has a nondegenerate maximum eigenvalue $E \le 1$,
and the corresponding right ($e_{i}$) and left ($\hat{e}_{i}$) 
eigenvectors can be chosen as 
$e_{i} \succ 0$ and $\hat{e}_{i} \succ 0$~\cite{Franosch02}.
The MCT critical point is characterized by
$E^{c} = 1$.
Let $e_{i}$ and $\hat{e}_{i}$ specifically 
denote the right and left eigenvectors,
respectively, of the stability matrix $C_{ij}^{c}$ at the
critical point:
$\sum_{j} C_{ij}^{c} e_{j} = e_{i}$,
$\sum_{i} \hat{e}_{i} C_{ij}^{c} = \hat{e}_{j}$.
The eigenvectors are fixed uniquely by requiring
$\sum_{i} \hat{e}_{i} e_{i} = 1$ and
$\sum_{i} \hat{e}_{i} \,
[{\bf e}({\bf S}-{\bf F}^{c}){\bf e}]_{i} = 1$.
These eigenvectors can be used evaluate the critical amplitudes
\begin{equation}
{\bf H}_{q} = [{\bf S}_{q} - {\bf F}_{q}^{c}] \, {\bf e}_{q} \,
[{\bf S}_{q} - {\bf F}_{q}^{c}] \, ,
\end{equation}
and the exponent parameter $\lambda$
\begin{equation}
\lambda = \sum_{q} \sum_{a,b}
\hat{e}_{q}^{ab} \, {\cal F}_{q}^{ab}[{\bf H}] \, .
\end{equation}
The above formulation for molecular systems is essentially the same as 
the one for simple systems, the only difference being the appearance of
matrices in place of scalar quantities.
It is then obvious that all the universal results concerning the
MCT-liquid-glass-transition dynamics,
as originally derived for simple systems~\cite{Goetze91b}, 
are valid also for MCT for molecular systems 
with such replacement of scalar quantities with matrices. 

Since the LW OTP molecule consists of three sites, 
a natural choice would be $n = 3$, 
and the MCT equations (\ref{eq:GLE-v}) and (\ref{eq:MCT-v})
for this model are given by $3 \times 3$ matrix equations.
On the other hand, the analysis of the simulation results presented in
Sec.~\ref{sec:MD-4} concerning the unusual feature at intermediate 
$q \approx q_{\rm GC}$ 
suggests the importance of taking into
account the spatial correlation of GC through
the static structure factor $S_{q}^{\rm GC}$. 
This implies that, in accounting for the unusual feature
at $q \approx q_{\rm GC}$, 
it might be necessary to include GC
as the additional 4th site in the theoretical calculations. 
This leads to another formulation for LW OTP 
based on $4 \times 4$-matrix site-site density correlators 
$F_{q}^{ab}(t)$, for which 
the MCT equations (\ref{eq:GLE-v}) and (\ref{eq:MCT-v})
are given by $4 \times 4$ matrix equations.
The required static inputs $S_{q}^{ab}$ for $a,b=1, \cdots, 4$
in the new formulation 
can also be determined from MD simulations. 
By comparing theoretical results with and without
including GC, 
one can judge whether the unusual feature
at the intermediate wavenumbers
$q \approx q_{\rm GC}$ discussed in Sec.~\ref{sec:MD-4}
stems from the coupling to the GC dynamics or not. 

One might think that the inclusion of GC as an additional site
is an ad hoc procedure, which is motivated in view
of the simulation results.
This kind of problems can occur in the site-density formulation
since the site-density fluctuations $\rho_{\vec q}^{a}$
defined for {\em finite} number of sites
do not provide a complete set of variables
describing the dynamics of molecules. 
This is in contrast to the tensor-density fluctuations
as adopted in Refs.~\cite{Winkler00,Fabbian99b,Theis00,Schilling97,Theis98,Letz00,Theenhaus01,Franosch97c,Goetze00c},
which do provide a complete set of variables for molecules and also 
naturally incorporate the GC correlation. 
However, as mentioned in Sec.~\ref{sec:MD-1},
the most important density fluctuations relevant for
the structural slowing down manifest themselves as the
main peak at $q = q_{\rm max}$ in $S_{q}^{\rm N}$ (the cage effect).
Such correlations are incorporated in the site-density
formulation even with the natural choice for the number of sites
($n=3$ for LW OTP).
Indeed, it will be shown below that the MCT equations with $n=3$
can describe the basic feature of the simulation results,
and this formulation is already useful. 
Our attitude here to include GC as an additional site (resulting in $n=4$)
is only for the investigation of 
the dynamical features at $q \approx q_{\rm GC}$.
In fact, we will see that the inclusion of GC does not alter
significantly the results for the wavenumbers other than
$q \approx q_{\rm GC}$. 

\subsection{Density dependence of 
$\mbox{\boldmath $T_{c}$}$ and 
$\mbox{\boldmath $\lambda$}$}
\label{sec:MCT-2}

\begin{figure}
\includegraphics[width=0.65\linewidth]{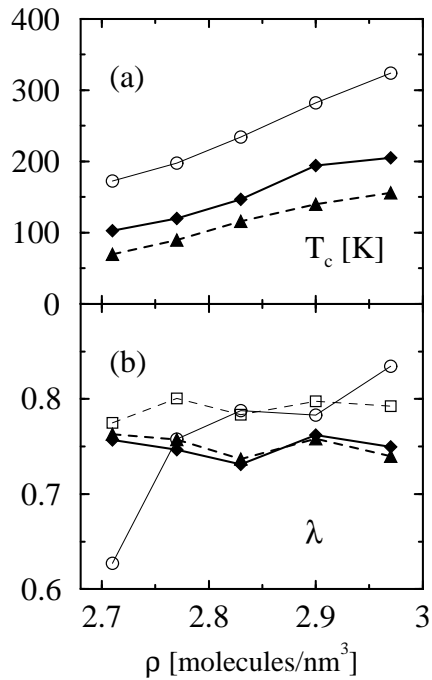}
\caption{
Same as in Fig.~\protect\ref{fig:Tc-lambda-MD},
but here theoretical results are included as well.
In both (a) and (b),
MCT results with and without including GC are denoted
as filled diamonds (connected with thick solid line) and 
filled triangles (connected with thick dashed line), respectively.}
\label{fig:Tc-lambda-comparison}
\end{figure}

Figures~\ref{fig:Tc-lambda-comparison}(a) and (b) respectively show
the theoretical results for $T_{c}$ and $\lambda$ 
as a function of the density $\rho$,
along with the corresponding simulation results to be denoted
as $T_{c}^{\rm MD}$ and $\lambda^{\rm MD}$ from here on in this section. 
It is seen from Fig.~\ref{fig:Tc-lambda-comparison}(a) that
both of the theoretical $T_{c}$, with and without including GC, 
exhibit qualitatively the same density dependence as that of $T_{c}^{\rm MD}$.
Concerning $\lambda$ shown in Fig.~\ref{fig:Tc-lambda-comparison}(b),
the theoretical results with and without including GC 
are practically the same, and are nearly independent of $\rho$. 
The latter feature is in accord with the simulation result
shown as squares in Fig.~\ref{fig:Tc-lambda-comparison}(b).
The agreement between 
the theoretical and simulation results for the value of $\lambda$ 
is reasonable in view of the error bars in 
estimating $\lambda^{\rm MD}$
as discussed in connection with Fig.~\ref{fig:Tc-lambda-MD}(b). 

The theoretical $T_{c}$ with including GC is in better agreement
with $T_{c}^{\rm MD}$, but is still located considerably below 
$T_{c}^{\rm MD}$ at all the densities investigated.
One might think that the discrepancy between the theoretical $T_{c}$ 
and $T_{c}^{\rm MD}$
is much larger than the one known for the hard-sphere system (HSS): 
MCT for HSS yields the critical packing fraction $\varphi_{c}$
which differs only about 7\% from the experimental value $\varphi_{c}^{\rm exp}$
for hard-sphere colloids~\cite{Foffi03}, 
for which HSS is known to serve as a good model.
(The previous estimate of the difference between the theoretical 
$\varphi_{c}$ and the experimental $\varphi_{c}^{\rm exp}$ was
about 15\%~\cite{Goetze91b}, but a recent analysis performed in
Ref.~\cite{Foffi03} indicates that half of this error is due 
to the use of the Percus-Yevick static structure factor as input
instead of simulated one.)
However, we argue in the following that
the discrepancy between the theoretical $T_{c}$ and $T_{c}^{\rm MD}$ for LW OTP
is of comparable size to the one for HSS. 

To this end, we notice that the $\rho$ dependence of $T_{c}^{\rm MD}$
could be condensed into a nearly constant effective coupling parameter
$\Gamma_{c}^{\rm MD} = 1.57 \pm 0.05$ ({\em cf.} Sec.~\ref{sec:MD-2}).
We found that, to a reasonable extent, this holds also for the
theoretical results:
from the $\rho$ dependence of the theoretical $T_{c}$, 
one gets 
$\Gamma_{c} = 1.77 \pm 0.07$ and $1.9 \pm 0.1$ with and without
including GC, respectively.
Thus, in terms of the critical effective coupling parameter $\Gamma_{c}$, 
the discrepancies between the theoretical and simulation results are only
within 13\% and 20\% with and without including GC, 
and are of comparable size to the discrepancy found for HSS. 
A similar analysis has been done for binary mixture of LJ
particles (BMLJ)~\cite{Nauroth97},
for which it was discussed that the difference between
$T_{c} = 0.922$ from MCT and 
$T_{c}^{\rm MD} = 0.435$ from the MD simulation 
is comparable ($\approx 20$ \%) to the one for HSS
when quantified in terms of the effective coupling parameter. 
We thus conclude that our theoretical estimate of $T_{c}$ for LW OTP is
within the comparable error bar as the one for HSS and BMLJ. 

However, we notice that our theoretical results exhibit an unconventional
feature in that $T_{c}$ 
is {\em underestimated} compared to $T_{c}^{\rm MD}$.
This is in contrast to all previous MCT calculations, 
where MCT is found to overestimate $T_{c}$ 
(or underestimate $\varphi_{c}$ when the packing fraction is concerned):
e.g., 
$\varphi_{c} = 0.546 < \varphi_{c}^{\rm exp} \approx 0.58$ for 
HSS~\cite{Foffi03}, and 
$T_{c} = 0.922 > T_{c}^{\rm MD} = 0.435$ for BMLJ~\cite{Nauroth97}.
In particular, this is also in contrast to MCT results for molecules
based on the tensor-density 
formulation~\cite{Winkler00,Fabbian99b,Theis00}.
Concerning this problem, we notice that 
including the triple direct correlation function $c_{3}$ 
was found to move up $T_{c}$ for LW OTP
more than by a factor of 2~\cite{Rinaldi01}.
Although the finding in Ref.~\cite{Rinaldi01} is based on 
the simplified scalar MCT equations dealing with the correlators
$\phi_{q}^{\rm N}(t)$ only, 
we expect that the inclusion of $c_{3}$ would significantly move up
$T_{c}$ shown in Fig.~\ref{fig:Tc-lambda-comparison}(a)
and would lead to $T_{c} > T_{c}^{\rm MD}$ in agreement with
the previous MCT studies.

Such theoretical calculations with including $c_{3}$, however, shall not
be performed in the present work. 
This is basically because 
an accurate evaluation of $c_{3}$ from MD simulations
is quite a demanding task.
Furthermore, it was also found in Ref.~\cite{Rinaldi01} for LW OTP
that including $c_{3}$ does not significantly alter theoretical results 
other than $T_{c}$, 
and we consider that MCT without $c_{3}$ captures the essential physics
in the dynamics of supercooled LW OTP. 
Indeed, as we will see below, 
our MCT results without the use of $c_{3}$ are 
in semiquantitative agreement with the simulation results. 

\subsection{Critical nonergodicity parameters and critical amplitudes}
\label{sec:MCT-3}

\begin{figure}
\includegraphics[width=0.65\linewidth]{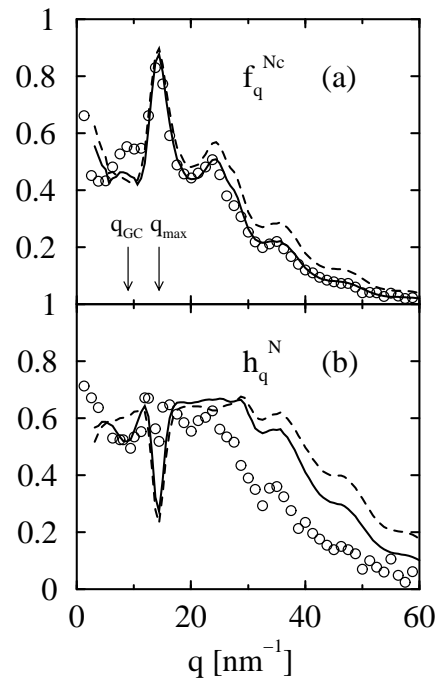}
\caption{
Comparison of MD simulation and theoretical results for
(a) the critical nonergodicity parameters $f_{q}^{{\rm N}c}$ and
(b) the critical amplitudes $h_{q}^{\rm N}$ 
of the correlators $\phi_{q}^{\rm N}(t)$
for the density $\rho = 2.71$ molecules$/$nm$^{3}$.
In both (a) and (b), circles denote the simulation results
as determined from the fit according to 
Eq.~(\protect\ref{eq:vs}), while 
solid and dashed lines refer to MCT results with 
and without including GC, respectively.
The simulation results for $h_{q}^{\rm N}$ are in arbitrary
units.}
\label{fig:fqc-NN-comparison}
\end{figure}

Figures~\ref{fig:fqc-NN-comparison}(a) and (b) compare
the theoretical and MD simulation results for
the critical nonergodicity parameters $f_{q}^{{\rm N}c}$
and the critical amplitudes $h_{q}^{\rm N}$ 
of the correlators $\phi_{q}^{\rm N}(t)$, respectively.
In both of these figures, 
solid and dashed lines respectively 
refer to the MCT results with and without including GC,
while circles denote the simulation results. 

Concerning $f_{q}^{{\rm N}c}$ shown in 
Fig.~\ref{fig:fqc-NN-comparison}(a), it is seen that,
even without including GC, the $q$ dependence of the
simulation result is well reproduced by the theory, 
especially for wavenumbers $q \gtrsim q_{\rm max}$. 
However, the theory without GC fails to reproduce
the peak in $f_{q}^{{\rm N}c}$ at $q \approx q_{\rm GC}$ 
found in the simulation result.
This peak is reproduced by the theory which includes GC,
although its magnitude is underestimated. 
Furthermore, the overall agreement with the simulation result
becomes better also for $q \gtrsim q_{\rm max}$ 
by including GC.

Also for $h_{q}^{\rm N}$ exhibited in 
Fig.~\ref{fig:fqc-NN-comparison}(b), 
the overall $q$ dependence of the simulation result
is well reproduced even by the theory without GC. 
On the other hand, the minimum in $h_{q}^{\rm N}$ at
$q \approx q_{\rm GC}$ found in the simulation result
is accounted for only by the theory which includes GC. 
The found improvement in the theoretical results for
$f_{q}^{{\rm N}c}$ and $h_{q}^{\rm N}$ at $q \approx q_{\rm GC}$,
which is achieved by including GC,
supports the idea that the unusual wavenumber dependence 
discussed in Sec.~\ref{sec:MD-4} 
is basically due to the coupling to the GC dynamics. 
A further support shall be discussed in Sec.~\ref{sec:MCT-4}
from the analysis of the $\alpha$-relaxation times. 

\subsection{Dynamics in the 
\protect$\mbox{\boldmath$\alpha$}$-relaxation region}
\label{sec:MCT-4}

As discussed in Ref.~\cite{Chong03} for LW OTP based on the 
MD simulation, 
the most faithful tests of MCT concerning the dynamics
should be performed in the $\alpha$-relaxation part 
starting from the plateau regime. 
This is because the approach toward the plateau, for which
MCT predicts an asymptotic power-law $\sim t^{-a}$ 
$(0 < a < 0.5)$~\cite{Goetze91b},
was found to be almost completely masked by the microscopic 
dynamics. 
In view of this, tests of the theoretical results for density
correlators shall be done 
in terms of MCT $\alpha$-master curves. 
Of particular relevance here is the MCT second scaling law --
also referred to as the superposition principle --
which states that correlators in the
$\alpha$-relaxation region for different
temperatures can be superposed on top of each other
simply by rescaling the time scale: 
\begin{equation}
\phi_{q}^{\rm X}(t) = \tilde{\phi}_{q}^{\rm X}(t/\tau_{q}^{\rm X}) \, .
\label{eq:superposition}
\end{equation}
Here $\tilde{\phi}_{q}^{\rm X}(\tilde{t})$ denotes 
the $\alpha$-master function.

In the strict test of the MCT second scaling law, 
on the other hand, 
one cannot freely choose the scale $\tau_{q}^{\rm X}$,
which can depend on the choice of the variable ${\rm X}$
as well as on the wavenumber $q$. 
According to MCT, there exists a single time scale, say
$\tau$, characterizing the $\alpha$ relaxation of all the
correlators~\cite{Goetze91b}. 
Thus, instead of Eq.~(\ref{eq:superposition}), 
one actually has for the MCT second scaling law 
\begin{equation}
\phi_{q}^{\rm X}(t) = \tilde{\phi}_{q}^{\rm X}(t/\tau) \, .
\label{eq:strict-superposition}
\end{equation}
The MCT $\alpha$-master function 
$\tilde{\phi}_{q}^{\rm X}(\tilde{t})$
can be evaluated from the 
MCT equations at $T = T_{c}$ up to an overall time scale, 
with the initial behavior given by the von Schweidler 
law, Eq.~(\ref{eq:vs})~\cite{Goetze91b}. 

\begin{figure}
\includegraphics[width=0.75\linewidth]{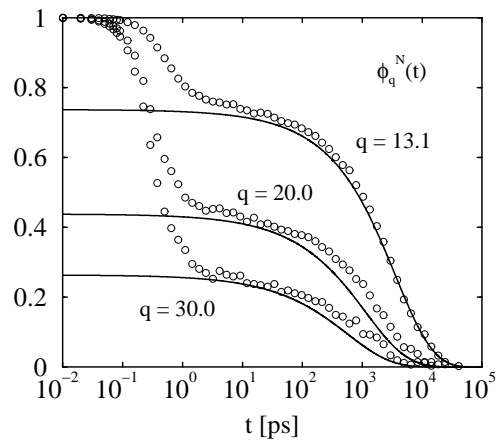}
\caption{
Comparison of MD simulation and theoretical results for
the correlators $\phi_{q}^{\rm N}(t)$ 
for the density $\rho = 2.71$ molecules$/$nm$^{3}$
at three wavenumbers $q = 13.1$, 20.0, and 30.0 nm$^{-1}$.
Circles denote the simulation results at $T = 190$ K.
Solid lines refer to the $\alpha$-master curves 
$\tilde{\phi}_{q}^{\rm N}(t/\tau)$ from MCT including GC,
where $\tau$ has been chosen so that both 
the simulation and theoretical 
curves at $q = 13.1$ nm$^{-1}$
yield the same $\alpha$-relaxation time $\tau_{q}^{\rm N}$ 
when fitted with Eq.~(\ref{eq:Kohlrausch}).}
\label{fig:NN-master-curve}
\end{figure}

The mentioned second scaling law of MCT implies that 
the test of the MCT $\alpha$-master functions
against simulation results for density correlators 
should be done by adjusting the single time scale $\tau$ only.
Such test is performed in 
Fig.~\ref{fig:NN-master-curve} 
for the correlators $\phi_{q}^{\rm N}(t)$ 
at three wavenumbers 
$q = 13.1$, 20.0, and 30.0 nm$^{-1}$, 
which are close to the first peak, the first minimum,
and the second minimum in $S_{q}^{\rm N}$, respectively 
({\em cf.} Fig.~\ref{fig:static-MD}). 
In Fig.~\ref{fig:NN-master-curve}, circles refer to the simulation results
for $\phi_{q}^{\rm N}(t)$
at $\rho = 2.71$ molecules$/$nm$^{3}$ and $T = 190$ K.
Solid lines denote the $\alpha$-master curves 
$\tilde{\phi}_{q}^{\rm N}(t/\tau)$ from MCT which includes GC,
where $\tau$ has been chosen so that both the theoretical and simulation
curves at $q = 13.1$ nm$^{-1}$
yield the same $\alpha$-relaxation time $\tau_{q}^{\rm N}$
when fitted with Eq.~(\ref{eq:Kohlrausch}).
It is seen from Fig.~\ref{fig:NN-master-curve} that the
MCT $\alpha$-master curves describe well the time dependence
of the simulation results for $\phi_{q}^{\rm N}(t)$ in the
$\alpha$-relaxation region, including the
relative $\alpha$-relaxation times for
different wavenumbers. 
(Notice that, from the construction of $\tau$, the real
test of the relative $\alpha$-relaxation times is performed
only for $q = 20.0$ and 30.0 nm$^{-1}$ in 
Fig.~\ref{fig:NN-master-curve}.)
In particular, 
the relaxation stretching, which is 
pronounced for $q = 20.0$ and 30.0 nm$^{-1}$, is well reproduced
by the theory. 
Indeed, from the Kohlrausch-law fit according to Eq.~(\ref{eq:Kohlrausch}), 
we found, e.g., for $q = 20.0$ nm$^{-1}$, 
$\beta_{q}^{\rm N} = 0.62$ from the MCT $\alpha$-master function and
0.60 from the simulation curve. 
A similar quantitative test of MCT against simulation results for 
density correlators, which uses the single time scale 
as an adjustable parameter, 
has been performed in
Ref.~\cite{Kob02} for binary mixture of LJ particles. 

\begin{figure}
\includegraphics[width=0.65\linewidth]{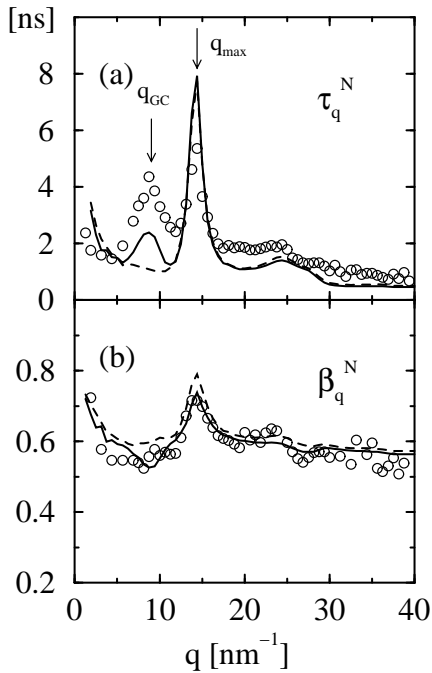}
\caption{
Comparison of MD simulation and theoretical results for
(a) the $\alpha$-relaxation times $\tau_{q}^{\rm N}$
and 
(b) the stretching exponents $\beta_{q}^{\rm N}$
of the correlators $\phi_{q}^{\rm N}(t)$
for the density $\rho = 2.71$ molecules$/$nm$^{3}$. 
In both (a) and (b), 
circles denote the simulations results at $T = 190$ K, 
while solid and dashed lines refer to results 
from the MCT $\alpha$-master curves
with and without
including GC, respectively.
The overall time scale of the theoretical results 
for $\tau_{q}^{\rm N}$
has been chosen so as to reproduce the same $\tau_{q}^{\rm N}$
at $q = 13.1$ nm$^{-1}$ as that from the MD simulation.}
\label{fig:NN-tau-alpha-beta-comparison}
\end{figure}

Theoretical and simulation results for 
the $\alpha$-relaxation times $\tau_{q}^{\rm N}$ 
and the stretching exponents $\beta_{q}^{\rm N}$ 
of the correlators $\phi_{q}^{\rm N}(t)$ for the whole wavenumbers
are compared in Figs.~\ref{fig:NN-tau-alpha-beta-comparison}(a) and (b),
respectively,
which are obtained from the fits according to Eq.~(\ref{eq:Kohlrausch}). 
In these figures, 
solid and dashed lines respectively
denote the MCT results with and without including GC,
whereas circles refer to the simulation results.
Again, the overall scale of the theoretical results for the 
$\alpha$-relaxation times has been chosen so as to reproduce the same
$\tau_{q}^{\rm N}$ at $q = 13.1$ nm$^{-1}$
as that from the MD simulation. 
It is seen that, even without including GC, the 
$q$ dependence of the simulation results for $\tau_{q}^{\rm N}$ and 
$\beta_{q}^{\rm N}$ is well reproduced by the theory at 
the semiquantitative level,
especially for the wavenumbers $q \gtrsim q_{\rm max}$. 
For $q < q_{\rm max}$, on the other hand, 
it is seen from 
Fig.~\ref{fig:NN-tau-alpha-beta-comparison}(a) that 
the unusual peak in the $\alpha$-relaxation times
at intermediate $q \approx q_{\rm GC}$, 
as observed in the simulation result, is reproduced only by the
theory which includes GC, 
although its magnitude is still underestimated.
This again supports the idea that the unusual peak is 
basically due to the coupling to the GC dynamics. 

\section{Summary and concluding remarks}
\label{sec:conclusions}

In this paper, we reported MD simulation results performed for
a model of molecular liquid OTP developed by Lewis and Wahnstr\"om,
paying special attention to the wavenumber dependence of 
the structural $\alpha$ relaxation of the collective dynamics,  
and showed that 
the simulation results for the model share many features 
with experimental data for real system (Sec.~\ref{sec:MD}). 
We then demonstrated that theoretical results based on MCT
captures the simulation results at the semiquantitative level 
(Sec.~\ref{sec:MCT}):
it is found that MCT yields a fair estimate of the critical 
temperature $T_{c}$ and the exponent parameter $\lambda$
including their density dependence, and predicts
the wavenumber dependence of dynamical quantities rather well,
in particular near the first sharp diffraction peak $q_{\rm max}$
of the static structure factor $S_{q}^{\rm N}$.
Through these investigations, 
we established the relevance of our
theoretical results and their interpretation in understanding 
experimental data for real system. 

As described in Sec.~\ref{sec:MD-1}, 
major intermolecular correlations manifest themselves as the peaks
at $q = q_{\rm max}$ in $S_{q}^{\rm N}$ and 
at $q = q_{\rm GC}$ in $S_{q}^{\rm GC}$.
On the other hand, the most pronounced temperature dependence 
in the static structure factors
shows up at $q = q_{\rm max}$ in $S_{q}^{\rm N}$ 
({\em cf.} the inset of Fig.~\ref{fig:static-MD}(a)). 
This indicates that the structural slowing down and
anomalous glassy features in the dynamics 
upon lowering $T$ are primarily caused by the 
intermolecular correlation 
manifested as the main peak in $S_{q}^{\rm N}$ (the cage effect).
This is supported by the observation in Sec.~\ref{sec:MCT}
that the theoretical results,
which do not take into account $S_{q}^{\rm GC}$, already capture
the basic features of the simulation results.

On the other hand, though of subordinate nature in the above sense,
the simulation results for LW OTP exhibit interesting and unusual
properties at intermediate wavenumbers $q \approx q_{\rm GC}$,
which reflect purely dynamical effects. 
As discussed in Sec.~\ref{sec:MD-5}, 
similar features can also be observed 
in experimental data for real system.
We argued that such unusual features
for $q \approx q_{\rm GC}$ are basically due to the
coupling to the GC dynamics.
This is because, compared to the simulation results, 
the theoretical results for $q \approx q_{\rm GC}$ were found to be
improved by including the spatial correlation of GC through $S_{q}^{\rm GC}$.
However, 
there still remain quantitative discrepancies
between the theoretical and simulation results for $q \approx q_{\rm GC}$
compared with the agreement found for the other wavenumber regime. 
This implies that the present theory still lacks some
features which might be relevant for the dynamics at intermediate
wavenumbers $q \approx q_{\rm GC}$.
Now, we provide additional evidence showing that this is the case.

\begin{figure}
\includegraphics[width=0.65\linewidth]{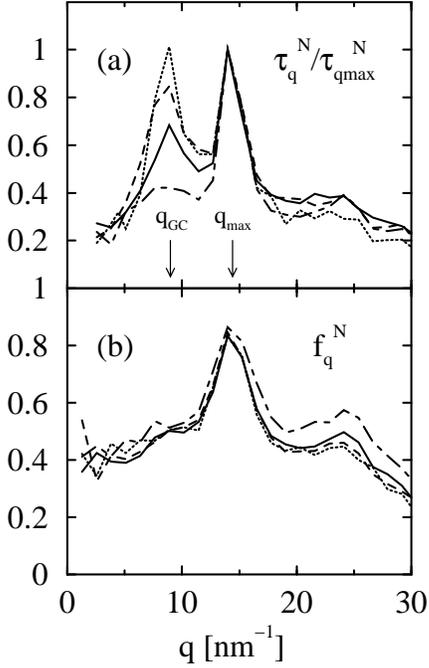}
\caption{
MD simulation results for 
(a) the rescaled $\alpha$-relaxation times
$\tau_{q}^{\rm N} / \tau_{q_{\rm max}}^{\rm N}$ and
(b) the plateau heights $f_{q}^{\rm N}$ 
of the correlators $\phi_{q}^{\rm N}(t)$ for the
density $\rho = 2.83$ molecules$/$nm$^{3}$.
The results at
$T = 260$ K (solid lines), 300 K (dashed lines), and 
340 K (dotted lines) refer to $T > T_{c} \approx 234$ K
({\em cf.} Fig.~\protect\ref{fig:Tc-lambda-MD}), 
whereas those at $T = 230$ K (dash-dotted lines)
to $T < T_{c}$.}
\label{fig:OTP-tau-N-vs-T}
\end{figure}

Figure~\ref{fig:OTP-tau-N-vs-T}(a) exhibits the
simulation result for the temperature
dependence of the $\alpha$-relaxation times normalized 
by the one at $q = q_{\rm max}$,
$\tau_{q}^{\rm N}/\tau_{q_{\rm max}}^{\rm N}$,
of the density correlators $\phi_{q}^{\rm N}(t)$ for the density
$\rho = 2.83$ molecules$/$nm$^{3}$.
MCT predicts the universal asymptotic power-law increase of the
$\alpha$-relaxation times~\cite{Goetze91b}.
This implies that the ratio of the $\alpha$-relaxation times
like the one shown in Fig.~\ref{fig:OTP-tau-N-vs-T}(a)
should be temperature independent for $T$ close to but above 
$T_{c}$,
which is referred to as the $\alpha$-relaxation-scale coupling.
However, Fig.~\ref{fig:OTP-tau-N-vs-T}(a) clearly demonstrates
that this universal prediction of MCT is violated around (and only around)
the wavenumber $q_{\rm GC}$. 
(For the other wavenumbers, 
it is seen that the $\alpha$-relaxation-scale coupling holds 
fairly well, including $T = 230$ K which is below 
$T_{c} \approx 234$ K for the density considered.)
Thus, the found temperature dependence of the
ratio $\tau_{q}^{\rm N}/\tau_{q_{\rm max}}^{\rm N}$
around $q = q_{\rm GC}$
is beyond the implication of MCT,
and the present theory, formulated within the framework of MCT,
cannot be used to explain such an finding. 
Unfortunately, 
due to the presence of incoherent background, 
it is not clear whether experimental results for 
the coherent $\alpha$-relaxation times reported in 
Ref.~\cite{Toelle98} also exhibit such a temperature 
dependence at $q \approx q_{\rm GC}$. 
(See also below for some other related 
simulation and experimental studies.) 

Combining the results shown in 
Figs.~\ref{fig:fqc-NN-MD}(a), \ref{fig:fqc-NN-MD}(b) and 
\ref{fig:OTP-tau-N-vs-T}(a),
one recognizes another interesting property of the unusual
peak at $q \approx q_{\rm GC}$ that it is more pronounced 
at {\em lower} density and at {\em higher} temperature.
In view of the fact that the overall shape of the LW OTP molecule
is well described as a sphere of the van der Waals radius 
$r_{\rm W} = 0.37$ nm ({\em cf}. Fig.~\ref{fig:picture-OTP}),
the dynamics at low-density, high-temperature regime
is expected to be dominated by the spatial correlation
of molecule's geometrical center, i.e., by $S_{q}^{\rm GC}$. 
This is in contrast to the high-density, low-temperature glassy
regime where the dynamics is primarily determined by the
cage effect manifested as the main peak in $S_{q}^{\rm N}$.
Thus, one might conjecture that the unusual feature at
$q \approx q_{\rm GC}$ is an inheritance from the low-density,
high-temperature dynamics, 
and that this cannot be explained by the universal
predictions of MCT 
since, up to the lowest temperature investigated, 
the dynamics at $q \approx q_{\rm GC}$ might not have yet reached
the asymptotic regime for which those MCT predictions are applicable. 
Such a possibility has been discussed in 
Refs.~\cite{Goetze00c,Chong02b},
where the standard MCT scenario for the glass-transition dynamics
was shown to be modified for some reorientational correlators
due to precursor phenomena of a nearby type-A transition.

The above conjecture, however, might not be appropriate
in the present case in view of the following result.
Figure~\ref{fig:OTP-tau-N-vs-T}(b) exhibits 
the temperature dependence of the plateau heights $f_{q}^{\rm N}$ 
of the correlators $\phi_{q}^{\rm N}(t)$.
For $T > T_{c}$, the plateau height can be obtained
from the von-Schweidler-law fit according to Eq.~(\ref{eq:vs}).
Although Eq.~(\ref{eq:vs}) cannot be employed for fitting 
the correlators referring to $T < T_{c}$,
we used it just to estimate the
plateau heights $f_{q}^{\rm N}$ for $T = 230$ K. 
(For $T = 230$ K, the correlators $\phi_{q}^{\rm N}(t)$ exhibit
well developed plateau regime as, e.g., shown in 
Fig.~3 of Ref.~\cite{Chong03}, and the plateau heights can easily
be estimated from such graphs.
We confirmed that the so-estimated plateau heights are in good
agreement with the ones based on the ad-hoc use of the
von-Schweidler-law fit.)

Since the plateau heights for $T > T_{c}$ corresponds to the
critical nonergodicity parameters $f_{q}^{{\rm N}c}$
({\em cf.} Eq.~(\ref{eq:vs})),
MCT predicts that they should be temperature independent. 
Furthermore, MCT predicts for $T < T_{c}$ the 
increase of the plateau height, 
$f_{q}^{N} - f_{q}^{{\rm N}c} \propto h_{q}^{\rm N} > 0$.
The results shown in Fig.~\ref{fig:OTP-tau-N-vs-T}(b) 
are, within the statistical errors,
consistent with these universal predictions of MCT.
Notice that this holds also for the intermediate wavenumbers
$q \approx q_{\rm GC}$. 
Thus, the plateau heights around $q = q_{\rm GC}$ have already
reached the asymptotic regime for which the universal
MCT description is adequate.
Since the plateau heights also quantify the strengths
of the $\alpha$-relaxation processes, 
it is then difficult to imagine that only the $\alpha$-relaxation
times have not yet reached the MCT asymptotic regime,
and the above conjecture introduced to explain the
finding in Fig.~\ref{fig:OTP-tau-N-vs-T}(a) 
might not be appropriate. 

The position $q \approx q_{\rm GC}$ where the unusual feature
we discussed occurs in LW OTP is compatible with the inverse of its van der
Waals radius ({\em cf}. Sec.~\ref{sec:MD-1}), 
i.e., it is connected to the overall size of the molecule. 
It is interesting to note that similar unusual peaks were
found in a model for 
polymer around the intermediate wavenumber $q = 2 \pi / R_{g}$,
where $R_{g}$ denotes the radius of gyration~\cite{Aichele01-all}:
in this wavenumber regime, 
a shoulder is discernible in the critical nonergodicity parameters $f_{q}^{c}$,
and peaks are observable in the inverse of the critical amplitudes $1/h_{q}$,
the $\alpha$-relaxation times $\tau_{q}$, and the stretching exponents $\beta_{q}$
of the correlators which correspond to $\phi_{q}^{\rm N}(t)$
in the present paper. 
In particular, the ratio 
$\tau_{q}/\tau_{q_{\rm max}}$ of the $\alpha$-relaxation times
at $q \approx 2 \pi / R_{g}$ for this model 
exhibits the same temperature dependence
as the one shown in Fig.~\ref{fig:OTP-tau-N-vs-T}(a). 
A similar temperature dependence of the ratio $\tau_{q}/\tau_{q_{\rm max}}$
at intermediate $q$ range ($\approx 0.4 \, q_{\rm max}$)
was also found in the coherent neutron scattering
results for a real polymer system~\cite{Farago02}.
Thus, further investigations are necessary 
for a comprehensive understanding of the unusual features
at intermediate wavenumbers 
as observed in simulation and experimental results for
molecular and polymer systems. 

\begin{acknowledgments}

We thank H.~Z.~Cummins, W.~Kob, and W.~G\"otze for discussions and
suggestions. 
We acknowledge support from MIUR COFIN and FIRB
and from INFM PRA-GENFDT. 

\end{acknowledgments}


\begin{thebibliography}{54}
\expandafter\ifx\csname natexlab\endcsname\relax\def\natexlab#1{#1}\fi
\expandafter\ifx\csname bibnamefont\endcsname\relax
  \def\bibnamefont#1{#1}\fi
\expandafter\ifx\csname bibfnamefont\endcsname\relax
  \def\bibfnamefont#1{#1}\fi
\expandafter\ifx\csname citenamefont\endcsname\relax
  \def\citenamefont#1{#1}\fi
\expandafter\ifx\csname url\endcsname\relax
  \def\url#1{\texttt{#1}}\fi
\expandafter\ifx\csname urlprefix\endcsname\relax\def\urlprefix{URL }\fi
\providecommand{\bibinfo}[2]{#2}
\providecommand{\eprint}[2][]{\url{#2}}

\bibitem[{\citenamefont{G{\"o}tze}(1991)}]{Goetze91b}
\bibinfo{author}{\bibfnamefont{W.}~\bibnamefont{G{\"o}tze}}, in
  \emph{\bibinfo{booktitle}{Liquids, Freezing and Glass Transition}}, edited by
  \bibinfo{editor}{\bibfnamefont{J.-P.} \bibnamefont{Hansen}},
  \bibinfo{editor}{\bibfnamefont{D.}~\bibnamefont{Levesque}}, \bibnamefont{and}
  \bibinfo{editor}{\bibfnamefont{J.}~\bibnamefont{Zinn-Justin}}
  (\bibinfo{publisher}{North-Holland}, \bibinfo{address}{Amsterdam},
  \bibinfo{year}{1991}), p. \bibinfo{pages}{287}.

\bibitem[{\citenamefont{G{\"o}tze and Sj{\"o}gren}(1992)}]{Goetze92}
\bibinfo{author}{\bibfnamefont{W.}~\bibnamefont{G{\"o}tze}} \bibnamefont{and}
  \bibinfo{author}{\bibfnamefont{L.}~\bibnamefont{Sj{\"o}gren}},
  \bibinfo{journal}{Rep. Prog. Phys.} \textbf{\bibinfo{volume}{55}},
  \bibinfo{pages}{241} (\bibinfo{year}{1992}).

\bibitem[{\citenamefont{G{\"o}tze}(1999)}]{Goetze99}
\bibinfo{author}{\bibfnamefont{W.}~\bibnamefont{G{\"o}tze}},
  \bibinfo{journal}{J. Phys.: Condensed Matter} \textbf{\bibinfo{volume}{11}},
  \bibinfo{pages}{A1} (\bibinfo{year}{1999}).

\bibitem[{\citenamefont{T{\"o}lle et~al.}(1997)\citenamefont{T{\"o}lle,
  Schober, Wuttke, and Fujara}}]{Toelle97}
\bibinfo{author}{\bibfnamefont{A.}~\bibnamefont{T{\"o}lle}},
  \bibinfo{author}{\bibfnamefont{H.}~\bibnamefont{Schober}},
  \bibinfo{author}{\bibfnamefont{J.}~\bibnamefont{Wuttke}}, \bibnamefont{and}
  \bibinfo{author}{\bibfnamefont{F.}~\bibnamefont{Fujara}},
  \bibinfo{journal}{Phys. Rev. E} \textbf{\bibinfo{volume}{56}},
  \bibinfo{pages}{809} (\bibinfo{year}{1997}).

\bibitem[{\citenamefont{T{\"o}lle}(2001)}]{Toelle01}
\bibinfo{author}{\bibfnamefont{A.}~\bibnamefont{T{\"o}lle}},
  \bibinfo{journal}{Rep. Prog. Phys.} \textbf{\bibinfo{volume}{64}},
  \bibinfo{pages}{1473} (\bibinfo{year}{2001}).

\bibitem[{\citenamefont{Monaco et~al.}(2001)\citenamefont{Monaco, Fioretto,
  Comez, and Ruocco}}]{Monaco01}
\bibinfo{author}{\bibfnamefont{G.}~\bibnamefont{Monaco}},
  \bibinfo{author}{\bibfnamefont{D.}~\bibnamefont{Fioretto}},
  \bibinfo{author}{\bibfnamefont{L.}~\bibnamefont{Comez}}, \bibnamefont{and}
  \bibinfo{author}{\bibfnamefont{G.}~\bibnamefont{Ruocco}},
  \bibinfo{journal}{Phys. Rev. E} \textbf{\bibinfo{volume}{63}},
  \bibinfo{pages}{061502} (\bibinfo{year}{2001}).

\bibitem[{\citenamefont{Petry et~al.}(1991)\citenamefont{Petry, Bartsch,
  Fujara, Kiebel, Sillescu, and Farago}}]{Petry91}
\bibinfo{author}{\bibfnamefont{W.}~\bibnamefont{Petry}},
  \bibinfo{author}{\bibfnamefont{E.}~\bibnamefont{Bartsch}},
  \bibinfo{author}{\bibfnamefont{F.}~\bibnamefont{Fujara}},
  \bibinfo{author}{\bibfnamefont{M.}~\bibnamefont{Kiebel}},
  \bibinfo{author}{\bibfnamefont{H.}~\bibnamefont{Sillescu}}, \bibnamefont{and}
  \bibinfo{author}{\bibfnamefont{B.}~\bibnamefont{Farago}},
  \bibinfo{journal}{Z. Phys. B} \textbf{\bibinfo{volume}{83}},
  \bibinfo{pages}{175} (\bibinfo{year}{1991}).

\bibitem[{\citenamefont{Kiebel et~al.}(1992)\citenamefont{Kiebel, Bartsch,
  Debus, Fujara, Petry, and Sillescu}}]{Kiebel92}
\bibinfo{author}{\bibfnamefont{M.}~\bibnamefont{Kiebel}},
  \bibinfo{author}{\bibfnamefont{E.}~\bibnamefont{Bartsch}},
  \bibinfo{author}{\bibfnamefont{O.}~\bibnamefont{Debus}},
  \bibinfo{author}{\bibfnamefont{F.}~\bibnamefont{Fujara}},
  \bibinfo{author}{\bibfnamefont{W.}~\bibnamefont{Petry}}, \bibnamefont{and}
  \bibinfo{author}{\bibfnamefont{H.}~\bibnamefont{Sillescu}},
  \bibinfo{journal}{Phys. Rev. B} \textbf{\bibinfo{volume}{45}},
  \bibinfo{pages}{10301} (\bibinfo{year}{1992}).

\bibitem[{\citenamefont{Wuttke et~al.}(1993)\citenamefont{Wuttke, Kiebel,
  Bartsch, Fujara, Petry, and Sillescu}}]{Wuttke93}
\bibinfo{author}{\bibfnamefont{J.}~\bibnamefont{Wuttke}},
  \bibinfo{author}{\bibfnamefont{M.}~\bibnamefont{Kiebel}},
  \bibinfo{author}{\bibfnamefont{E.}~\bibnamefont{Bartsch}},
  \bibinfo{author}{\bibfnamefont{F.}~\bibnamefont{Fujara}},
  \bibinfo{author}{\bibfnamefont{W.}~\bibnamefont{Petry}}, \bibnamefont{and}
  \bibinfo{author}{\bibfnamefont{H.}~\bibnamefont{Sillescu}},
  \bibinfo{journal}{Z. Phys. B} \textbf{\bibinfo{volume}{91}},
  \bibinfo{pages}{357} (\bibinfo{year}{1993}).

\bibitem[{\citenamefont{Bartsch et~al.}(1995)\citenamefont{Bartsch, Fujara,
  Legrand, Petry, Sillescu, and Wuttke}}]{Bartsch95}
\bibinfo{author}{\bibfnamefont{E.}~\bibnamefont{Bartsch}},
  \bibinfo{author}{\bibfnamefont{F.}~\bibnamefont{Fujara}},
  \bibinfo{author}{\bibfnamefont{J.~F.} \bibnamefont{Legrand}},
  \bibinfo{author}{\bibfnamefont{W.}~\bibnamefont{Petry}},
  \bibinfo{author}{\bibfnamefont{H.}~\bibnamefont{Sillescu}}, \bibnamefont{and}
  \bibinfo{author}{\bibfnamefont{J.}~\bibnamefont{Wuttke}},
  \bibinfo{journal}{Phys. Rev. E} \textbf{\bibinfo{volume}{52}},
  \bibinfo{pages}{738} (\bibinfo{year}{1995}).

\bibitem[{\citenamefont{T{\"o}lle
  et~al.}(1998{\natexlab{a}})\citenamefont{T{\"o}lle, Schober, Wuttke, Randl,
  and Fujara}}]{Toelle98b}
\bibinfo{author}{\bibfnamefont{A.}~\bibnamefont{T{\"o}lle}},
  \bibinfo{author}{\bibfnamefont{H.}~\bibnamefont{Schober}},
  \bibinfo{author}{\bibfnamefont{J.}~\bibnamefont{Wuttke}},
  \bibinfo{author}{\bibfnamefont{O.~G.} \bibnamefont{Randl}}, \bibnamefont{and}
  \bibinfo{author}{\bibfnamefont{F.}~\bibnamefont{Fujara}},
  \bibinfo{journal}{Phys. Rev. Lett.} \textbf{\bibinfo{volume}{80}},
  \bibinfo{pages}{2374} (\bibinfo{year}{1998}{\natexlab{a}}).

\bibitem[{\citenamefont{T{\"o}lle
  et~al.}(1998{\natexlab{b}})\citenamefont{T{\"o}lle, Wuttke, Schober, Randl,
  and Fujara}}]{Toelle98}
\bibinfo{author}{\bibfnamefont{A.}~\bibnamefont{T{\"o}lle}},
  \bibinfo{author}{\bibfnamefont{J.}~\bibnamefont{Wuttke}},
  \bibinfo{author}{\bibfnamefont{H.}~\bibnamefont{Schober}},
  \bibinfo{author}{\bibfnamefont{O.~G.} \bibnamefont{Randl}}, \bibnamefont{and}
  \bibinfo{author}{\bibfnamefont{F.}~\bibnamefont{Fujara}},
  \bibinfo{journal}{Eur. Phys. J. B} \textbf{\bibinfo{volume}{5}},
  \bibinfo{pages}{231} (\bibinfo{year}{1998}{\natexlab{b}}).

\bibitem[{\citenamefont{G{\"o}tze and Sj{\"o}gren}(1991)}]{Goetze91}
\bibinfo{author}{\bibfnamefont{W.}~\bibnamefont{G{\"o}tze}} \bibnamefont{and}
  \bibinfo{author}{\bibfnamefont{L.}~\bibnamefont{Sj{\"o}gren}},
  \bibinfo{journal}{Phys. Rev. A} \textbf{\bibinfo{volume}{43}},
  \bibinfo{pages}{5442} (\bibinfo{year}{1991}).

\bibitem[{Fof()}]{Foffi03}
\bibinfo{note}{G. Foffi, W. G\"otze, F. Sciortino, P. Tartaglia, and Th.
  Voigtmann, e-print cond-mat/0309007.}

\bibitem[{\citenamefont{Nauroth and Kob}(1997)}]{Nauroth97}
\bibinfo{author}{\bibfnamefont{M.}~\bibnamefont{Nauroth}} \bibnamefont{and}
  \bibinfo{author}{\bibfnamefont{W.}~\bibnamefont{Kob}},
  \bibinfo{journal}{Phys. Rev. E} \textbf{\bibinfo{volume}{55}},
  \bibinfo{pages}{657} (\bibinfo{year}{1997}).

\bibitem[{\citenamefont{Sciortino and Kob}(2001)}]{Sciortino01}
\bibinfo{author}{\bibfnamefont{F.}~\bibnamefont{Sciortino}} \bibnamefont{and}
  \bibinfo{author}{\bibfnamefont{W.}~\bibnamefont{Kob}},
  \bibinfo{journal}{Phys. Rev. Lett.} \textbf{\bibinfo{volume}{86}},
  \bibinfo{pages}{648} (\bibinfo{year}{2001}).

\bibitem[{\citenamefont{Winkler et~al.}(2000)\citenamefont{Winkler, Latz,
  Schilling, and Theis}}]{Winkler00}
\bibinfo{author}{\bibfnamefont{A.}~\bibnamefont{Winkler}},
  \bibinfo{author}{\bibfnamefont{A.}~\bibnamefont{Latz}},
  \bibinfo{author}{\bibfnamefont{R.}~\bibnamefont{Schilling}},
  \bibnamefont{and} \bibinfo{author}{\bibfnamefont{C.}~\bibnamefont{Theis}},
  \bibinfo{journal}{Phys. Rev. E} \textbf{\bibinfo{volume}{62}},
  \bibinfo{pages}{8004} (\bibinfo{year}{2000}).

\bibitem[{\citenamefont{Fabbian et~al.}(1999)\citenamefont{Fabbian, Latz,
  Schilling, Sciortino, Tartaglia, and Theis}}]{Fabbian99b}
\bibinfo{author}{\bibfnamefont{L.}~\bibnamefont{Fabbian}},
  \bibinfo{author}{\bibfnamefont{A.}~\bibnamefont{Latz}},
  \bibinfo{author}{\bibfnamefont{R.}~\bibnamefont{Schilling}},
  \bibinfo{author}{\bibfnamefont{F.}~\bibnamefont{Sciortino}},
  \bibinfo{author}{\bibfnamefont{P.}~\bibnamefont{Tartaglia}},
  \bibnamefont{and} \bibinfo{author}{\bibfnamefont{C.}~\bibnamefont{Theis}},
  \bibinfo{journal}{Phys. Rev. E} \textbf{\bibinfo{volume}{60}},
  \bibinfo{pages}{5768} (\bibinfo{year}{1999}).

\bibitem[{\citenamefont{Theis et~al.}(2000)\citenamefont{Theis, Sciortino,
  Latz, Schilling, and Tartaglia}}]{Theis00}
\bibinfo{author}{\bibfnamefont{C.}~\bibnamefont{Theis}},
  \bibinfo{author}{\bibfnamefont{F.}~\bibnamefont{Sciortino}},
  \bibinfo{author}{\bibfnamefont{A.}~\bibnamefont{Latz}},
  \bibinfo{author}{\bibfnamefont{R.}~\bibnamefont{Schilling}},
  \bibnamefont{and}
  \bibinfo{author}{\bibfnamefont{P.}~\bibnamefont{Tartaglia}},
  \bibinfo{journal}{Phys. Rev. E} \textbf{\bibinfo{volume}{62}},
  \bibinfo{pages}{1856} (\bibinfo{year}{2000}).

\bibitem[{\citenamefont{Lewis and Wahnstr{\"o}m}(1994)}]{Lewis94}
\bibinfo{author}{\bibfnamefont{L.~J.} \bibnamefont{Lewis}} \bibnamefont{and}
  \bibinfo{author}{\bibfnamefont{G.}~\bibnamefont{Wahnstr{\"o}m}},
  \bibinfo{journal}{Phys. Rev. E} \textbf{\bibinfo{volume}{50}},
  \bibinfo{pages}{3865} (\bibinfo{year}{1994}).

\bibitem[{\citenamefont{Rinaldi et~al.}(2001)\citenamefont{Rinaldi, Sciortino,
  and Tartaglia}}]{Rinaldi01}
\bibinfo{author}{\bibfnamefont{A.}~\bibnamefont{Rinaldi}},
  \bibinfo{author}{\bibfnamefont{F.}~\bibnamefont{Sciortino}},
  \bibnamefont{and}
  \bibinfo{author}{\bibfnamefont{P.}~\bibnamefont{Tartaglia}},
  \bibinfo{journal}{Phys. Rev. E} \textbf{\bibinfo{volume}{63}},
  \bibinfo{pages}{061210} (\bibinfo{year}{2001}).

\bibitem[{\citenamefont{Mossa et~al.}(2002)\citenamefont{Mossa, {E. La Nave},
  Stanley, Donati, Sciortino, and Tartaglia}}]{Mossa02}
\bibinfo{author}{\bibfnamefont{S.}~\bibnamefont{Mossa}},
  \bibinfo{author}{\bibnamefont{{E. La Nave}}},
  \bibinfo{author}{\bibfnamefont{H.~E.} \bibnamefont{Stanley}},
  \bibinfo{author}{\bibfnamefont{C.}~\bibnamefont{Donati}},
  \bibinfo{author}{\bibfnamefont{F.}~\bibnamefont{Sciortino}},
  \bibnamefont{and}
  \bibinfo{author}{\bibfnamefont{P.}~\bibnamefont{Tartaglia}},
  \bibinfo{journal}{Phys. Rev. E} \textbf{\bibinfo{volume}{65}},
  \bibinfo{pages}{041205} (\bibinfo{year}{2002}).

\bibitem[{\citenamefont{Chong and Sciortino}(2003)}]{Chong03}
\bibinfo{author}{\bibfnamefont{S.-H.} \bibnamefont{Chong}} \bibnamefont{and}
  \bibinfo{author}{\bibfnamefont{F.}~\bibnamefont{Sciortino}},
  \bibinfo{journal}{Europhys. Lett.} \textbf{\bibinfo{volume}{64}},
  \bibinfo{pages}{197} (\bibinfo{year}{2003}).

\bibitem[{\citenamefont{Schilling and Scheidsteger}(1997)}]{Schilling97}
\bibinfo{author}{\bibfnamefont{R.}~\bibnamefont{Schilling}} \bibnamefont{and}
  \bibinfo{author}{\bibfnamefont{T.}~\bibnamefont{Scheidsteger}},
  \bibinfo{journal}{Phys. Rev. E} \textbf{\bibinfo{volume}{56}},
  \bibinfo{pages}{2932} (\bibinfo{year}{1997}).

\bibitem[{\citenamefont{Theis and Schilling}(1998)}]{Theis98}
\bibinfo{author}{\bibfnamefont{C.}~\bibnamefont{Theis}} \bibnamefont{and}
  \bibinfo{author}{\bibfnamefont{R.}~\bibnamefont{Schilling}},
  \bibinfo{journal}{J. Non-Cryst. Solids} \textbf{\bibinfo{volume}{235--237}},
  \bibinfo{pages}{106} (\bibinfo{year}{1998}).

\bibitem[{\citenamefont{Letz et~al.}(2000)\citenamefont{Letz, Schilling, and
  Latz}}]{Letz00}
\bibinfo{author}{\bibfnamefont{M.}~\bibnamefont{Letz}},
  \bibinfo{author}{\bibfnamefont{R.}~\bibnamefont{Schilling}},
  \bibnamefont{and} \bibinfo{author}{\bibfnamefont{A.}~\bibnamefont{Latz}},
  \bibinfo{journal}{Phys. Rev. E} \textbf{\bibinfo{volume}{62}},
  \bibinfo{pages}{5173} (\bibinfo{year}{2000}).

\bibitem[{\citenamefont{Theenhaus et~al.}(2001)\citenamefont{Theenhaus,
  Schilling, Latz, and Letz}}]{Theenhaus01}
\bibinfo{author}{\bibfnamefont{T.}~\bibnamefont{Theenhaus}},
  \bibinfo{author}{\bibfnamefont{R.}~\bibnamefont{Schilling}},
  \bibinfo{author}{\bibfnamefont{A.}~\bibnamefont{Latz}}, \bibnamefont{and}
  \bibinfo{author}{\bibfnamefont{M.}~\bibnamefont{Letz}},
  \bibinfo{journal}{Phys. Rev. E} \textbf{\bibinfo{volume}{64}},
  \bibinfo{pages}{051505} (\bibinfo{year}{2001}).

\bibitem[{\citenamefont{Franosch
  et~al.}(1997{\natexlab{a}})\citenamefont{Franosch, Fuchs, G{\"o}tze, Mayr,
  and Singh}}]{Franosch97c}
\bibinfo{author}{\bibfnamefont{T.}~\bibnamefont{Franosch}},
  \bibinfo{author}{\bibfnamefont{M.}~\bibnamefont{Fuchs}},
  \bibinfo{author}{\bibfnamefont{W.}~\bibnamefont{G{\"o}tze}},
  \bibinfo{author}{\bibfnamefont{M.~R.} \bibnamefont{Mayr}}, \bibnamefont{and}
  \bibinfo{author}{\bibfnamefont{A.~P.} \bibnamefont{Singh}},
  \bibinfo{journal}{Phys. Rev. E} \textbf{\bibinfo{volume}{56}},
  \bibinfo{pages}{5659} (\bibinfo{year}{1997}{\natexlab{a}}).

\bibitem[{\citenamefont{G{\"o}tze et~al.}(2000)\citenamefont{G{\"o}tze, Singh,
  and $\mbox{Th}$. Voigtmann}}]{Goetze00c}
\bibinfo{author}{\bibfnamefont{W.}~\bibnamefont{G{\"o}tze}},
  \bibinfo{author}{\bibfnamefont{A.~P.} \bibnamefont{Singh}}, \bibnamefont{and}
  \bibinfo{author}{\bibnamefont{$\mbox{Th}$. Voigtmann}},
  \bibinfo{journal}{Phys. Rev. E} \textbf{\bibinfo{volume}{61}},
  \bibinfo{pages}{6934} (\bibinfo{year}{2000}).

\bibitem[{\citenamefont{Chong and Hirata}(1998)}]{Chong98b}
\bibinfo{author}{\bibfnamefont{S.-H.} \bibnamefont{Chong}} \bibnamefont{and}
  \bibinfo{author}{\bibfnamefont{F.}~\bibnamefont{Hirata}},
  \bibinfo{journal}{Phys. Rev. E} \textbf{\bibinfo{volume}{58}},
  \bibinfo{pages}{6188} (\bibinfo{year}{1998}).

\bibitem[{\citenamefont{Chong et~al.}(2001{\natexlab{a}})\citenamefont{Chong,
  G{\"o}tze, and Singh}}]{Chong01}
\bibinfo{author}{\bibfnamefont{S.-H.} \bibnamefont{Chong}},
  \bibinfo{author}{\bibfnamefont{W.}~\bibnamefont{G{\"o}tze}},
  \bibnamefont{and} \bibinfo{author}{\bibfnamefont{A.~P.} \bibnamefont{Singh}},
  \bibinfo{journal}{Phys. Rev. E} \textbf{\bibinfo{volume}{63}},
  \bibinfo{pages}{011206} (\bibinfo{year}{2001}{\natexlab{a}}).

\bibitem[{\citenamefont{Chong et~al.}(2001{\natexlab{b}})\citenamefont{Chong,
  G{\"o}tze, and Mayr}}]{Chong01b}
\bibinfo{author}{\bibfnamefont{S.-H.} \bibnamefont{Chong}},
  \bibinfo{author}{\bibfnamefont{W.}~\bibnamefont{G{\"o}tze}},
  \bibnamefont{and} \bibinfo{author}{\bibfnamefont{M.~R.} \bibnamefont{Mayr}},
  \bibinfo{journal}{Phys. Rev. E} \textbf{\bibinfo{volume}{64}},
  \bibinfo{pages}{011503} (\bibinfo{year}{2001}{\natexlab{b}}).

\bibitem[{\citenamefont{Chong and G{\"o}tze}(2002{\natexlab{a}})}]{Chong02}
\bibinfo{author}{\bibfnamefont{S.-H.} \bibnamefont{Chong}} \bibnamefont{and}
  \bibinfo{author}{\bibfnamefont{W.}~\bibnamefont{G{\"o}tze}},
  \bibinfo{journal}{Phys. Rev. E} \textbf{\bibinfo{volume}{65}},
  \bibinfo{pages}{041503} (\bibinfo{year}{2002}{\natexlab{a}}).

\bibitem[{\citenamefont{Chong and G{\"o}tze}(2002{\natexlab{b}})}]{Chong02b}
\bibinfo{author}{\bibfnamefont{S.-H.} \bibnamefont{Chong}} \bibnamefont{and}
  \bibinfo{author}{\bibfnamefont{W.}~\bibnamefont{G{\"o}tze}},
  \bibinfo{journal}{Phys. Rev. E} \textbf{\bibinfo{volume}{65}},
  \bibinfo{pages}{051201} (\bibinfo{year}{2002}{\natexlab{b}}).

\bibitem[{\citenamefont{Franosch
  et~al.}(1997{\natexlab{b}})\citenamefont{Franosch, Fuchs, G{\"o}tze, Mayr,
  and Singh}}]{Franosch97}
\bibinfo{author}{\bibfnamefont{T.}~\bibnamefont{Franosch}},
  \bibinfo{author}{\bibfnamefont{M.}~\bibnamefont{Fuchs}},
  \bibinfo{author}{\bibfnamefont{W.}~\bibnamefont{G{\"o}tze}},
  \bibinfo{author}{\bibfnamefont{M.~R.} \bibnamefont{Mayr}}, \bibnamefont{and}
  \bibinfo{author}{\bibfnamefont{A.~P.} \bibnamefont{Singh}},
  \bibinfo{journal}{Phys. Rev. E} \textbf{\bibinfo{volume}{55}},
  \bibinfo{pages}{7153} (\bibinfo{year}{1997}{\natexlab{b}}).

\bibitem[{\citenamefont{Fuchs}(1994)}]{Fuchs94}
\bibinfo{author}{\bibfnamefont{M.}~\bibnamefont{Fuchs}}, \bibinfo{journal}{J.
  Non-Cryst. Solids} \textbf{\bibinfo{volume}{172--174}}, \bibinfo{pages}{241}
  (\bibinfo{year}{1994}).

\bibitem[{\citenamefont{Bondi}(1964)}]{Bondi64}
\bibinfo{author}{\bibfnamefont{A.}~\bibnamefont{Bondi}}, \bibinfo{journal}{J.
  Phys. Chem.} \textbf{\bibinfo{volume}{68}}, \bibinfo{pages}{441}
  (\bibinfo{year}{1964}).

\bibitem[{\citenamefont{Hansen and McDonald}(1986)}]{Hansen86}
\bibinfo{author}{\bibfnamefont{J.-P.} \bibnamefont{Hansen}} \bibnamefont{and}
  \bibinfo{author}{\bibfnamefont{I.~R.} \bibnamefont{McDonald}},
  \emph{\bibinfo{title}{Theory of Simple Liquids}}
  (\bibinfo{publisher}{Academic Press}, \bibinfo{address}{London},
  \bibinfo{year}{1986}), \bibinfo{edition}{2nd} ed.

\bibitem[{\citenamefont{Bernu et~al.}(1987)\citenamefont{Bernu, Hansen,
  Hiwatari, and Pastore}}]{Bernu87}
\bibinfo{author}{\bibfnamefont{B.}~\bibnamefont{Bernu}},
  \bibinfo{author}{\bibfnamefont{J.-P.} \bibnamefont{Hansen}},
  \bibinfo{author}{\bibfnamefont{Y.}~\bibnamefont{Hiwatari}}, \bibnamefont{and}
  \bibinfo{author}{\bibfnamefont{G.}~\bibnamefont{Pastore}},
  \bibinfo{journal}{Phys. Rev. A} \textbf{\bibinfo{volume}{36}},
  \bibinfo{pages}{4891} (\bibinfo{year}{1987}).

\bibitem[{\citenamefont{Roux et~al.}(1989)\citenamefont{Roux, Barrat, and
  Hansen}}]{Roux89}
\bibinfo{author}{\bibfnamefont{J.~N.} \bibnamefont{Roux}},
  \bibinfo{author}{\bibfnamefont{J.-L.} \bibnamefont{Barrat}},
  \bibnamefont{and} \bibinfo{author}{\bibfnamefont{J.-P.}
  \bibnamefont{Hansen}}, \bibinfo{journal}{J. Phys.: Condens. Matter}
  \textbf{\bibinfo{volume}{1}}, \bibinfo{pages}{7171} (\bibinfo{year}{1989}).

\bibitem[{\citenamefont{Bennemann et~al.}(1999)\citenamefont{Bennemann, Paul,
  Baschnagel, and Binder}}]{Bennemann99c}
\bibinfo{author}{\bibfnamefont{C.}~\bibnamefont{Bennemann}},
  \bibinfo{author}{\bibfnamefont{W.}~\bibnamefont{Paul}},
  \bibinfo{author}{\bibfnamefont{J.}~\bibnamefont{Baschnagel}},
  \bibnamefont{and} \bibinfo{author}{\bibfnamefont{K.}~\bibnamefont{Binder}},
  \bibinfo{journal}{J. Phys.: Condens. Matter} \textbf{\bibinfo{volume}{11}},
  \bibinfo{pages}{2179} (\bibinfo{year}{1999}).

\bibitem[{\citenamefont{Fuchs et~al.}(1992)\citenamefont{Fuchs, Hofacker, and
  Latz}}]{Fuchs92b}
\bibinfo{author}{\bibfnamefont{M.}~\bibnamefont{Fuchs}},
  \bibinfo{author}{\bibfnamefont{I.}~\bibnamefont{Hofacker}}, \bibnamefont{and}
  \bibinfo{author}{\bibfnamefont{A.}~\bibnamefont{Latz}},
  \bibinfo{journal}{Phys. Rev. A} \textbf{\bibinfo{volume}{45}},
  \bibinfo{pages}{898} (\bibinfo{year}{1992}).

\bibitem[{\citenamefont{Kob and Andersen}(1995)}]{Kob95b}
\bibinfo{author}{\bibfnamefont{W.}~\bibnamefont{Kob}} \bibnamefont{and}
  \bibinfo{author}{\bibfnamefont{H.~C.} \bibnamefont{Andersen}},
  \bibinfo{journal}{Phys. Rev. E} \textbf{\bibinfo{volume}{52}},
  \bibinfo{pages}{4134} (\bibinfo{year}{1995}).

\bibitem[{\citenamefont{Horbach and Kob}(2001)}]{Horbach01}
\bibinfo{author}{\bibfnamefont{J.}~\bibnamefont{Horbach}} \bibnamefont{and}
  \bibinfo{author}{\bibfnamefont{W.}~\bibnamefont{Kob}},
  \bibinfo{journal}{Phys. Rev. E} \textbf{\bibinfo{volume}{64}},
  \bibinfo{pages}{041503} (\bibinfo{year}{2001}).

\bibitem[{\citenamefont{Sciortino et~al.}(1997)\citenamefont{Sciortino,
  Fabbian, Chen, and Tartaglia}}]{Sciortino97}
\bibinfo{author}{\bibfnamefont{F.}~\bibnamefont{Sciortino}},
  \bibinfo{author}{\bibfnamefont{L.}~\bibnamefont{Fabbian}},
  \bibinfo{author}{\bibfnamefont{S.-H.} \bibnamefont{Chen}}, \bibnamefont{and}
  \bibinfo{author}{\bibfnamefont{P.}~\bibnamefont{Tartaglia}},
  \bibinfo{journal}{Phys. Rev. E} \textbf{\bibinfo{volume}{56}},
  \bibinfo{pages}{5397} (\bibinfo{year}{1997}).

\bibitem[{\citenamefont{Petry and Wuttke}(1995)}]{Petry95}
\bibinfo{author}{\bibfnamefont{W.}~\bibnamefont{Petry}} \bibnamefont{and}
  \bibinfo{author}{\bibfnamefont{J.}~\bibnamefont{Wuttke}},
  \bibinfo{journal}{Transp. Theory Stat. Phys.} \textbf{\bibinfo{volume}{24}},
  \bibinfo{pages}{1075} (\bibinfo{year}{1995}).

\bibitem[{\citenamefont{Kudchadkar and Wiest}(1995)}]{Kudchadkar95}
\bibinfo{author}{\bibfnamefont{S.~R.} \bibnamefont{Kudchadkar}}
  \bibnamefont{and} \bibinfo{author}{\bibfnamefont{J.~M.} \bibnamefont{Wiest}},
  \bibinfo{journal}{J. Chem. Phys.} \textbf{\bibinfo{volume}{103}},
  \bibinfo{pages}{8566} (\bibinfo{year}{1995}).

\bibitem[{Mos()}]{Mossa-OTP-all}
\bibinfo{note}{S. Mossa, R. {Di Leonardo}, G. Ruocco and M. Sampoli,
  Phys.~Rev.~E {\bf 62}, 612 (2000); S. Mossa, G. Ruocco and M. Sampoli,
  Phys.~Rev.~E {\bf 64}, 021511 (2001).}

\bibitem[{\citenamefont{Theis and Schilling}(1999)}]{Theis99}
\bibinfo{author}{\bibfnamefont{C.}~\bibnamefont{Theis}} \bibnamefont{and}
  \bibinfo{author}{\bibfnamefont{R.}~\bibnamefont{Schilling}},
  \bibinfo{journal}{Phys. Rev. E} \textbf{\bibinfo{volume}{60}},
  \bibinfo{pages}{740} (\bibinfo{year}{1999}).

\bibitem[{\citenamefont{Chong and Hirata}(1999)}]{Chong99}
\bibinfo{author}{\bibfnamefont{S.-H.} \bibnamefont{Chong}} \bibnamefont{and}
  \bibinfo{author}{\bibfnamefont{F.}~\bibnamefont{Hirata}},
  \bibinfo{journal}{J. Chem. Phys.} \textbf{\bibinfo{volume}{111}},
  \bibinfo{pages}{3083} (\bibinfo{year}{1999}).

\bibitem[{\citenamefont{Franosch and $\mbox{Th}$.
  Voigtmann}(2002)}]{Franosch02}
\bibinfo{author}{\bibfnamefont{T.}~\bibnamefont{Franosch}} \bibnamefont{and}
  \bibinfo{author}{\bibnamefont{$\mbox{Th}$. Voigtmann}}, \bibinfo{journal}{J.
  Stat. Phys.} \textbf{\bibinfo{volume}{109}}, \bibinfo{pages}{237}
  (\bibinfo{year}{2002}).

\bibitem[{\citenamefont{Kob et~al.}(2002)\citenamefont{Kob, Nauroth, and
  Sciortino}}]{Kob02}
\bibinfo{author}{\bibfnamefont{W.}~\bibnamefont{Kob}},
  \bibinfo{author}{\bibfnamefont{M.}~\bibnamefont{Nauroth}}, \bibnamefont{and}
  \bibinfo{author}{\bibfnamefont{F.}~\bibnamefont{Sciortino}},
  \bibinfo{journal}{J. Non-Cryst. Solids} \textbf{\bibinfo{volume}{307--310}},
  \bibinfo{pages}{181} (\bibinfo{year}{2002}).

\bibitem[{Aic()}]{Aichele01-all}
\bibinfo{note}{M.~Aichele and J.~Baschnagel, Eur.~Phys.~J.~E {\bf 5}, 229
  (2001); {\bf 5}, 245 (2001).}

\bibitem[{\citenamefont{Farago et~al.}(2002)\citenamefont{Farago, Arbe,
  Colmenero, Faust, Buchenau, and Richter}}]{Farago02}
\bibinfo{author}{\bibfnamefont{B.}~\bibnamefont{Farago}},
  \bibinfo{author}{\bibfnamefont{A.}~\bibnamefont{Arbe}},
  \bibinfo{author}{\bibfnamefont{J.}~\bibnamefont{Colmenero}},
  \bibinfo{author}{\bibfnamefont{R.}~\bibnamefont{Faust}},
  \bibinfo{author}{\bibfnamefont{U.}~\bibnamefont{Buchenau}}, \bibnamefont{and}
  \bibinfo{author}{\bibfnamefont{D.}~\bibnamefont{Richter}},
  \bibinfo{journal}{Phys. Rev. E} \textbf{\bibinfo{volume}{65}},
  \bibinfo{pages}{051803} (\bibinfo{year}{2002}).

\end{thebibliography}
\end{document}